\documentclass[journal]{IEEEtran}
\usepackage{algorithm}
\usepackage{algpseudocode}
\usepackage{amsfonts}
\usepackage{bm}
\usepackage{amsmath}
\usepackage{cite}
\usepackage{float}
\usepackage{amssymb}
\usepackage{enumerate}
\usepackage{color,xcolor}
\usepackage{graphicx}
\usepackage{multirow,tabularx}
\usepackage{subfigure}
\usepackage{hyperref}

\makeatletter

\newcommand{\Rmnum}[1]{\expandafter\@slowromancap\romannumeral #1@}
\makeatother

\usepackage{bookmark}

\usepackage{multicol}

\usepackage{flushend}

\graphicspath{{figure/}}

\ifCLASSINFOpdf

\else

\fi

\PassOptionsToPackage{draft}{hyperref}

\begin{document}

\title{{Capacity-Achieving MIMO-NOMA: Iterative LMMSE Detection}}

%

\author{\IEEEauthorblockN{Lei Liu, \emph{Member, IEEE,} Yuhao Chi, \emph{Member, IEEE,} Chau Yuen, \emph{Senior Member, IEEE,}\\ Yong Liang Guan, \emph{Senior Member, IEEE,} Ying Li, \emph{Member, IEEE}}
\thanks{Lei Liu was with the State Key Lab of Integrated Services Networks, Xidian University, Xi'an, 710071, China. He is now with the Department of Electronic Engineering, City University of Hong Kong, Hong Kong, SAR, China. (e-mail: lliu\_0@stu.xidian.edu.cn)}
\thanks{Yuhao Chi and Yong Liang Guan are with the School of Electrical and Electronic Engineering, Nanyang Technological University, Singapore (e-mail: chiyuhao1990@163.com, eylguan@ntu.edu.sg).}
\thanks{Chau Yuen and is with the Singapore University of Technology and Design, Singapore (e-mail: yuenchau@sutd.edu.sg)}
\thanks{Ying Li is with the State Key Lab of Integrated Services Networks, Xidian University, Xi'an, 710071, China (e-mail: yli@mail.xidian.edu.cn).}
}

\markboth{IEEE Transactions on Signal Processing}%
{Shell \MakeLowercase{\textit{et al.}}: Bare Demo of IEEEtran.cls for Journals}

\maketitle

\begin{abstract}
This paper considers a low-complexity iterative \emph{Linear Minimum Mean Square Error} (LMMSE) multi-user detector for the \emph{Multiple-Input and Multiple-Output} system with \emph{Non-Orthogonal Multiple Access} (MIMO-NOMA), {where multiple single-antenna users simultaneously communicate with a multiple-antenna base station (BS)}. While LMMSE being a linear detector has a low complexity, it has suboptimal performance in multi-user detection scenario due to the mismatch between LMMSE detection and multi-user decoding. Therefore, in this paper, we provide the matching conditions between the detector and decoders for MIMO-NOMA, which are then used to derive the achievable rate of the iterative detection. We prove that a matched iterative LMMSE detector can achieve \emph{(i)} the optimal capacity of symmetric MIMO-NOMA {with any number of users}, \emph{(ii)} the optimal sum capacity of asymmetric MIMO-NOMA {with any number of users}, \emph{(iii)} all the maximal extreme points in the capacity region of asymmetric MIMO-NOMA {with any number of users}, \emph{(iv)} all points in the capacity region of two-user and three-user asymmetric MIMO-NOMA systems. In addition, a kind of practical low-complexity error-correcting multiuser code, called irregular repeat-accumulate code, is designed to match the LMMSE detector. Numerical results shows that the bit error rate performance of the proposed iterative LMMSE detection outperforms the state-of-art methods and is within 0.8dB from the associated capacity limit.

\end{abstract}


\IEEEpeerreviewmaketitle
\section{Introduction}
{
Recent investigations have shown that \emph{Multi-user Multiple-Input Multiple-Output} (MU-MIMO), where multiple single-antenna users communicate with a multi-antenna \emph{Base Station} (BS), has become increasing important due to their potential applications in 5G cellular systems and beyond \cite{Argas2013, Rusek2013, biglieri2007, Marzetta2010, Lei2015, Lei2016}. In particular, massive MU-MIMO has been shown to be able to bring significant improvement in throughput and energy efficiency \cite{Rusek2013, Marzetta2010}.

Multiple access schemes, the fundamental techniques of coordinated multi-user communication in the physical layer, play the most important role in each cellular generation. \emph{Frequency Division Multiple Access} (FDMA), \emph{Time Division Multiple Access} (TDMA), \emph{Code Division Multiple Access} (CDMA), and \emph{Orthogonal Frequency-Division Multiple Access} (OFDMA) are the conventional \emph{Orthogonal Multiple Access} (OMA) schemes, which orthogonalize users in time/frequency/code domain to avoid multi-user interference \cite{tse2005, Dai_NOMA}. Due to the orthogonality of OMA, no inter-user interference exists at the receiver side. Hence, simple single-user signal processing in the conventional point-to-point communication can be directly used for OMA. However, there is no free lunch. First, OMA is not able to achieve all points in the capacity region of \emph{multiuser access channel} (MAC). Besides, massive connectivity will be the key scenario in the future wireless communication, and thus the limited radio resources cannot support the massive orthogonal access devices in the OMA any more. Apart from that, user scheduling such as resource allocation is required for orthogonal users in OMA, which leads to heavy additional overhead and results in large latency and high processing complexity in massive connectivity system.

Recently, \emph{Non-Orthogonal Multiple Access} (NOMA), where all the users can be served con-currently in the same time/frequency/code domain, has been identified as one of the key radio access technologies to increase the spectral efficiency and reduce latency in 5G mobile networks \cite{5GWhitepaper, METIS, Saito2013, Al-Imari2014, Dai_NOMA, Di2018, GL2018, Kim2015, Chen2015,LYW2017}. As opposed to OMA, the key concepts behind NOMA are summarized as follows \cite{Lei20161, Di2018, YC2018,LiPing2017,LYW2017}.
\begin{itemize}
\item All the users are allowed to be superimposed at the receiver in the same time/code/frequency domain.
\item All points in the capacity region of MAC are achievable.
\item Interference cancellation is performed at receiver, either \emph{Successive Interference Cancellation} (SIC) or \emph{Parallel Interference Cancellation} (PIC).
\end{itemize}

More recently, to enhance spectral efficiency and reduce latency, MIMO-NOMA that employs NOMA techniques over MU-MIMO is considered as a key air interface technology in the fifth-generation (5G) communication system \cite{Jiang2018,HW2018,LiPing2017,Ding2016,YC2018,Lei20161,LYW2017}. Therefore, we focus on MIMO-NOMA in this paper.
}
%

\subsection{Challenge of Multi-User Detection in MIMO-NOMA}Unlike the MIMO-OMA, signal processing in MIMO-NOMA will cost higher complexity and higher energy consumption at BS \cite{Rusek2013,biglieri2007}. Low-complexity uplink detection for MIMO-NOMA is a challenging problem due to the non-orthogonal interference between the users \cite{Rusek2013, Al-Imari2014, Kim2015, Chen2015}, especially when the number of users and the number of BS antennas are large. The optimal \emph{multiuser detector} (MUD) for the MIMO-NOMA, such as the \emph{maximum a-posteriori probability} (MAP) detector or \emph{maximum likelihood} (ML) detector, was proven to be an NP-hard and non-deterministic polynomial-time complete (NP-complete) problem \cite{Micciancio2001,verdu1984_1}. Furthermore, the complexity of optimal MUD grows exponentially with the number of users or the number of BS antennas, and polynomially with the size of signal constellation \cite{verdu1984_1,verdu1987}.

\subsection{Background of Low-Complexity Multi-User Detector}
Several low-complexity multi-user detectors have been proposed in the literature. They are mainly divided into three categories: uncoded detection, coded SIC detection,
and coded PIC detection.

\subsubsection{Uncoded Low-Complexity Detection}\label{uncoded_Det}Many low-complexity linear detections such as \emph{Matched Filter} (MF), \emph{Zero-Forcing} (ZF) receiver, \emph{Minimum Mean Square Error} (MMSE), and \emph{Message Passing Detector} (MPD) \cite{tse2005,Loeliger2006} are proposed for the practical systems. In addition, some iterative methods such as \emph{Jacobi method}, \emph{Richardson method}\cite{Axelsson1994, Bertsekas1989, Gao2014}, \emph{Belief Propagation} (BP) method, and iterative MPD \cite{ Lei2015,Lei2016,Lei20162, andrea2005} are put forward to further reduce the computational complexity by avoiding the unfavorable matrix inversion in the linear detections. Although being attractive from the complexity view point, these individual detectors are regarded to be sub-optimal MUDs, where decoding results are not fed back to the detector. As a result, the multi-user interference is not cancellated sufficiently.

\subsubsection{Coded SIC Detection}{SIC, where correct decoding results are fed back to the detector for perfect interference cancellation, is one of the key technologies to improve the detection performance. It is well known that for the MAC, the SIC is an optimal strategy and can achieve all points in the capacity region of MIMO-NOMA with time-sharing technology \cite{Cover2006, Gamal2012}.
Besides, the MMSE-SIC detector \cite{Wang1999,Studer2011} has been proposed to achieve the optimal performance \cite{tse2005}. Nevertheless, the following disadvantages make SIC infeasible when applying to the practical MIMO-NOMA \cite{Rusek2013,tse2005,verdu1998}.
\begin{itemize}
    \item The users are decoded one by one, which greatly increases the time delay.
    \item The decoding order is required to be known at both the transmitter and receiver, which results in additional overhead cost.
    \item It assumes that all the previous users' messages are recovered correctly and thus can be completely removed from the received signals. Nevertheless, in practice, the correct recovery is never be possible, which leads to error propagation during the interference cancellation.
    \item To achieve all points in the capacity region of MIMO-NOMA, time-sharing should be used, which needs cooperation between the users.
    \item The decoding order of SIC changes with the different channel state and different \emph{Quality of Service} (QoS), which brings a higher overhead cost.
\end{itemize}

\subsubsection{Coded PIC Detection} PIC, where users are parallelly recovered and messages exchanged between the detector and decoders are soft, is another promising technique for the practical MIMO-NOMA systems \cite{andrea2005, Gao2014,Lei2015,Wang1999}. This technique has been commonly used for the non-orthogonal MAC like the \emph{Code Division Multiple Access} (CDMA) systems \cite{tse2005,verdu1998} and the \emph{Interleave Division Multiple Access} (IDMA) systems \cite{Ping2003_1,Ping2004}. Various iterative detectors\footnote{For the uncoded iterative detector in Section I-B-1, the iteration is processed inside the detector. However, for the coded PIC detector, the iterative detection is performed between the detector and decoders, i.e., outside the detector.}, such as iterative \emph{Linear MMSE} (LMMSE) detector, iterative BP detector and iterative MPD \cite{Guo2008, Sanderovich2005, Yuan2014}. The advantages of iterative detection are listed as follows.
\begin{itemize}
    \item The complexity is very low, since the overall receiver is departed into many parallel low-complexity processors.
    \item Time delay is much lower than SIC, since the users are recovered in parallel.
    \item Error propagation is greatly mitigated, since user interference are cancellated in soft and thus perfect interference cancellation is not required.
    \item System overhead is reduced, since the preset decoding order is not required.
    \item User cooperation is removed, since time-sharing is not required.
\end{itemize}
 The existing PIC detections have a good simulative performance, but are regarded as suboptimal due to a performance gap to the associated capacity limit \cite{verdu1998}.  This is due to the fact that the detector and the decoders are designed separately and are not matched with each another, which results in performance loss although the decoding feedback is included for the detection.

\subsubsection{Principles of A Good Iterative Multi-User Detector}
From the review above, we conclude the key principles in designing a good iterative multi-user detector.
\begin{itemize}
    \item Multi-user interference cancellation and discrete signal reconstruction are performed respectively by MUD and user detectors.
    \item The decoding results should be fed back to the detector for a thorough interference cancellation.
    \item The detector and multi-user code should be jointly designed and matched with each other to avoid rate loss. In particular, the multi-user channel code should be optimized for the super-channel that encompasses the MIMO-NOMA channel and the multi-user detector.
\end{itemize}
The achievable rate analysis of such iterative detection for MIMO-NOMA is an intriguing problem.
{
\subsection{Relationship with Interference Channel and Vector Multiple Access Channel}
To clarify the relationship between interference channel (IC), vector multiple-access channel (VMAC) and MIMO-NOMA channel. We first give the definitions of IC and VMAC below.
\begin{itemize}
\item IC considers multiple transmitters and multiple receivers, and transmitter cooperation and receiver cooperation are not allowed (i.e. multiple scalar/vector inputs and multiple scalar/vector outputs).
\item VMAC considers  multiple transmitters and a single receiver, and both transmitters and receiver are equipped with multiple antennas (i.e. multiple vector inputs and a vector output).
\end{itemize}
Hence, the MIMO-NOMA channel (multiple scalar inputs and a vector output) discussed in this paper is different from IC because only a single receiver is considered. Moreover, the MIMO-NOMA channel is a special case of VMAC if each transmitter is only equipped with single antenna.

It is well known that the capacity of IC \cite{Han1981} is still an open issue. In addition, the capacity of general VMAC is only solved by a numerical algorithm \cite{YuW2004}. In contrast, MIMO-NOMA channel (or VMAC with single-antenna transmitters) has a closed-form capacity region, which has been solved in \cite{Han1979}, see also \cite{tse2005} and \cite{Gamal2012} for more details.

}

\subsection{Gap Between P2P MIMO and MIMO-NOMA}
The \emph{Extrinsic Information Transfer} (EXIT) \cite{Ashikhmin2004,Brink2001}, \emph{MSE-based Transfer Chart} (MSTC) \cite{Guo2005,Bhattad2007}, area theorem and matching theorem \cite{Ashikhmin2004, Brink2001, Bhattad2007, Guo2005} are the main methods to analyse the achievable rate or the \emph{Bit Error Rate} (BER) performance of MIMO systems. It is proven that a well-designed single-code with linear precoding and iterative LMMSE detection achieves the capacity of the MIMO systems \cite{Yuan2014}. However, this results only applies to \emph{point-to-point} (P2P) MIMO systems.

Since there is no user collaboration in MIMO-NOMA, the precoding in P2P MIMO \cite{Yuan2014} cannot be used. Besides, the singular value decomposition (SVD) and water-filling in \cite{Yuan2014} are unachievable in multi-user MIMO NOMA too, since there is no channel information at transmitters. Furthermore, only one user rate is analyzed in P2P MIMO \cite{Yuan2014}, but in MIMO-NOMA, the whole achievable rate region that contains all the user rates needs to be established. Apart from that, the non-orthogonal multi-user interference makes the problem be more complicated. For example, the decoding processes of the non-orthogonal users in MIMO-NOMA interfere with each other, which results in a much more complicated MSTC functions and area theorems. In summary, the results in P2P MIMO (e.g. \cite{Yuan2014}) cannot be cannot be straightforwardly applied to analyze the achievable rates of the iterative detection for MIMO-NOMA.}\vspace{-0.2cm}

\subsection{Contributions}
In this paper, the achievable rate analysis of the iterative LMMSE detection is provided for MIMO-NOMA, which shows that the low-complexity iterative LMMSE can be rate region optimal if it is properly designed. The contributions of this paper are summarized as follows\footnote{In points a, b, c and d, the ideal SCM codes (with infinite layers and infinite length), which are designed to match the SINR-variance transfer curves of LMMSE detection, are used for the multiuser codes.}.
\begin{enumerate}[a)]
\item Matching conditions and area theorems of the iterative detector are proposed for  MIMO-NOMA.

\item Achievable rate analysis of iterative LMMSE detection are provided.

\item Analytical proofs are derived for the designed iterative LMMSE detection to achieve:
\begin{itemize}
\item the capacity of symmetric MIMO-NOMA {with any number of users},
\item the sum capacity of the asymmetric MIMO-NOMA {with any number of users},
 \item all the maximal extreme points in the capacity region of the asymmetric MIMO-NOMA {with any number of users}, and
\item all points in the capacity region of two-user and three-user asymmetric MIMO-NOMA.
\end{itemize}
\item We prove that the elementary signal estimator (ESE) of IDMA in  Multiple Input and Signal Output (MISO) and the maximal ratio combiner (MRC) in  Multiple Output and Signal Input (SIMO) are two special cases of iterative LMMSE receiver. Hence, both ESE of IDMA in MISO and MRC in SIMO are sum capacity achieving.
\item An algorithm is provided to design a practical iterative LMMSE detection.

\item A kind of capacity-approaching multi-user NOMA code for the LMMSE detection, in the form of a special (non-standard) \emph{Irregular Repeat-Accumulate }(IRA) multiuser code, is systematically constructed. This special IRA multi-user code must be designed in conjunction with the LMMSE detection to produce extrinsic transfer functions that satisfy a certain constraint among the different users.

\item Numerical results show that our iterative LMMSE detection with optimized IRA code outperforms the existing methods, and is within 0.8dB from the associated capacity limit.
\end{enumerate}

{From the information theoretic point of view, to the best of our knowledge, this is the first work that proves that a proper designed PIC (joint design of the iterative LMMSE detection and the multi-user code)  can achieve the capacity of MIMO-NOMA with low complexity. From the practical point of view, the jointly designed iterative LMMSE detection (PIC) has significant improvement in the BER performances over the existing iterative receivers (including both SIC and PIC) in a variety of system loads.

\emph{Comments:} It is well known that finite-length coding will lead to rate loss. In this paper, when we refer to the proposed iterative LMMSE achieving the capacity (sum capacity or all points in the capacity region) of MIMO-NOMA, infinite-length channel codes are considered by default. Specifically, in this paper, we use an ideal SCM code (with infinite layers and infinite length), which is designed to match the SINR-variance transfer curves of LMMSE detection. The existence of such code is rigorously proved in APPENDIX \ref{APP:Code_existence}. }

This paper is organized as follows. In Section II, the MIMO-NOMA system and iterative LMMSE detection are introduced. The matching conditions and area theorems for the MIMO-NOMA are elaborated in Section III. Section IV provides the achievable rate analysis. Important properties and special cases of the iterative LMMSE detection are given in Section V. Practical multiuser code design is provided in Section VI. Numerical results are shown in Section VII.

\section{System Model and iterative LMMSE detection}
Consider an uplink MU-MIMO system that showed in Fig.~\ref{f2}: {$N_{u}$ autonomous single-antenna terminals simultaneously communicate with an array of $N_{r}$ antennas of the BS \cite{Marzetta2010,Rusek2013}. Here,  $N_{u}$ and $N_{r}$ can be any finite positive integers. Since all the users interfere with each other at the receiver and are non-orthogonal in the time, frequency and code domain, it is thus named MIMO-NOMA}\footnote{Here, MIMO-NOMA is different from IC since only a single receiver is considered. Moreover, MIMO-NOMA is also different from VMAC since each transmitter is only equipped with single antenna.}. The system is represented as
\begin{equation}\label{e1}
{\mathbf{y}_t}= \mathbf{H}{\mathbf{x}^{tr}(t)} + \mathbf{n}(t),\quad t\in \mathcal{N},\;\;\mathcal{N}=\{1,\cdots,N\}
\end{equation}
where $\mathbf{H}$ is an $N_{r} \times N_{u}$ channel matrix, $\mathbf{n}(t)\sim\mathcal{CN}^{{N_{r}}}(0,\sigma_n^2)$ an independent additive white Gaussian noise (AWGN), $\mathbf{x}^{tr}(t)=[x^{tr}_1(t),\cdots,x^{tr}_{N_u}(t)]^T$ the transmission, and $\mathbf{y}_t$ the received vector at time $t$. In this paper, we consider the block fading channel \cite{tse2005}, i.e., $\mathbf{H}$ is fixed during one block transmission and known at the BS. When the channel is block fading, in time-division duplexing (TDD) mode, it is possible for the BS to estimate the downlink channel when receiving message from the uplink. In frequency-division duplexing (FDD) mode, it is possible for the receiver feedback the channel to BS. However, as these are standard assumption for many others in the literature, we will not describe in details.

\subsection{Transmitters}
\begin{figure}[htb]
  \centering
  \includegraphics[width=9.5cm]{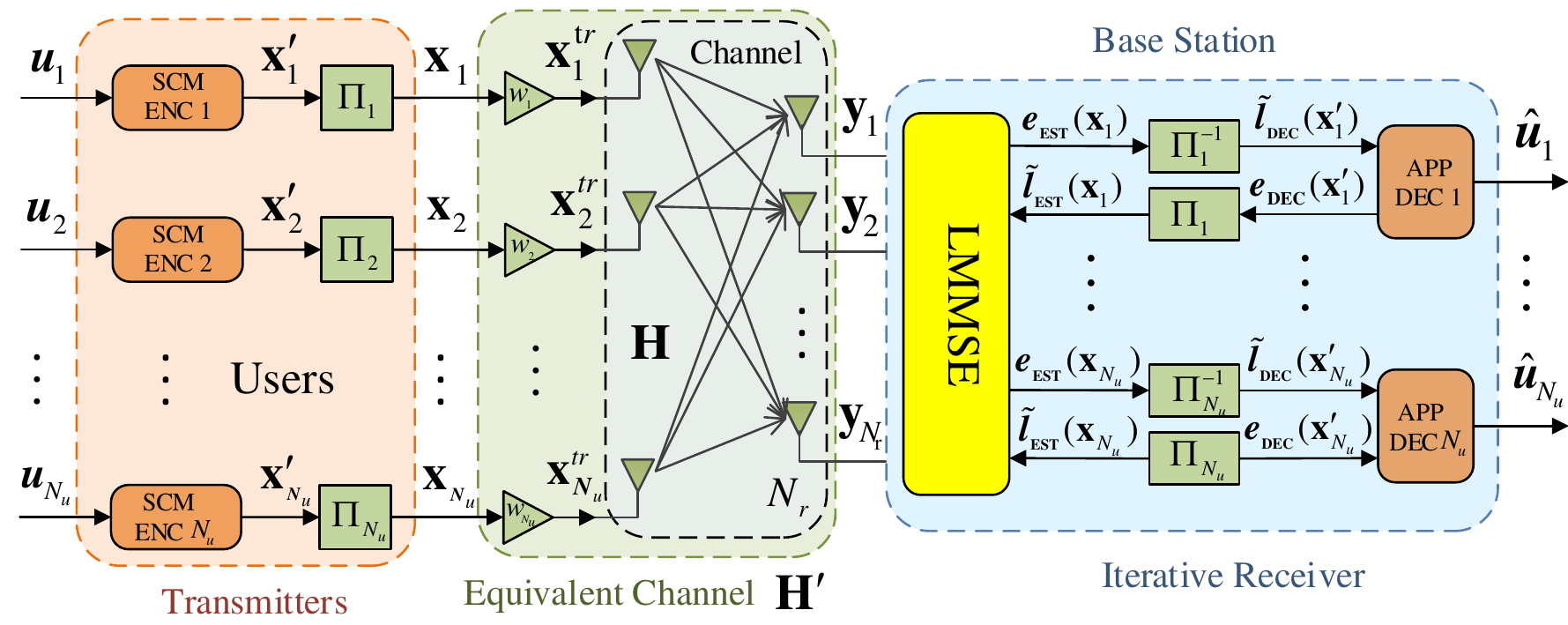}\\
  \caption{Block diagram of MIMO-NOMA system. SCM ENC is the superposition coded modulation encoder and APP DEC is the \emph{a-posteriori} probability decoder. $\Pi_i$ and $\Pi_i^{-1}$ denotes the interleaver and de-interleaver. LMMSE represents the LMMSE detector. The equivalent channel $\mathbf{H}'$ contains the channel $\mathbf{H}$ and the power parameter of each user $w_i$, $i\in\mathcal{N}_u$.}\label{f2}
\end{figure}

As illustrated in Fig. \ref{f2}, at user $i$ ($\mathop{i}\in \mathcal{N}_{u},\; \mathcal{N}_{u}= \left\{{{{1,2,}} \cdots {{,N_{u}}}} \right\}$), an information sequence ${\mathbf{u}}_i $ is encoded by an error-correcting code into an $N$-length sequence ${\mathbf{x}}'_{i}$, which is interleaved by an {$N$-length} independent random interleaver\footnote{The interleavers improve the system performance by enhancing the randomness of the messages or the channel noise, and avoiding the short cycles in the system factor graph \cite{Ping2003_1, Ping2004,Andrews2007}.} $\Pi_{i}$ to get $\mathbf{x}_{i}=[x_{i,1},x_{i,2},\cdots,x_{i,N}]^T$. We assume that each $x_{i,t}$ is taken over the points in a discrete signaling constellation $\mathcal{S}=\{s_1,s_2,\cdots,s_{|\mathcal{S}|}\}$. After that, the $\mathbf{x}_{i}$ is scaled with $w_i$, and we then get the transmitting $\mathbf{x}^{tr}_{i}$, $\mathop{i}\in \mathcal{N}_{u}$. Let $\sigma^2_{x_i}=1$ denote the normalized variance of $\mathbf{x}_{i}$, and $\mathbf{K}_{\mathbf{x}}$ be power constraint diagonal matrix whose diagonal elements are $w^2_i, \mathop{i}\in \mathcal{N}_{u}$. Therefore, the system can be rewritten to
\begin{equation}\label{e2}
{\mathbf{y}_t}=\mathbf{H}{\mathbf{K}_{\mathbf{x}}^{1/2}\mathbf{x}(t)} + \mathbf{n}(t)
= \mathbf{H}'{\mathbf{x}(t)} + \mathbf{n}(t),\quad t\in \mathcal{N},
\end{equation}
where ${\mathbf{x}(t)}=[x_1(t),\dots,x_{N_u}(t)]^T$.

\subsection{Capacity region of MIMO-NOMA}
 Let $\mathbf{Y}$ denote the received random vector, and $\mathbf{X}$ represent the transmitting random vector. Assuming $S\subseteq \mathcal{N}_u$, $S^c\subseteq \mathcal{N}_u/S$ and $S\cup S^c=\mathcal{N}_u$, the partial channel matrix is denoted as $\mathbf{H}'_S=[\{\mathbf{h}'_i, i\in S\}]_{N_r\times |S|}$, where $\mathbf{h}'_i$ is the $i$th column of $\mathbf{H}'$. Similar definition is applied to $\mathbf{X}_S$. Let $R_i$ be the rate of user $i$ and $R_S=\sum\limits_{i\in S}{R_i}$ represent the sum rate of the users in set $S$. Then, capacity region\footnote{Different from the interference channel whose capacity is still an open issue and the vector multiple-access channel whose capacity only has a numerical solution, the capacity calculation of MIMO-NOMA is trivial and has been has been well studied in \cite{Han1979, tse2005}.} $\mathbf{\mathcal{R}}_\mathcal{S}$ of the MIMO-NOMA system is given by \cite{Cover2006, Gamal2012}
\begin{eqnarray}\label{e17}
{R_S} \!\le\! I(\mathbf{Y};{\mathbf{X}_S}|{\mathbf{X}_{{S^c}}})
\!=\! \log|\mathbf{I}_{|\mathcal{S}|} \!+ \! \frac{1}{\sigma_n^2}\mathbf{H}'^H_S\mathbf{H}'_S|,
 \; \forall S \!\subseteq \!\mathcal{N}_u,
\end{eqnarray}
where $|\mathbf{A}|$ denotes the determinant of $\mathbf{A}$. The sum rate is
\begin{eqnarray}\label{e18}
R_{sum}= R_{\mathcal{N}_u}
= \log|\mathbf{I}_{N_u} + \frac{1}{\sigma_n^2}\mathbf{H}'^H\mathbf{H}'|.
\end{eqnarray}

\subsection{Iterative Receiver}
We adopt a joint detection-decoding iterative receiver, which is widely used in the multiple-access systems \cite{Lei20162,Wang1999,Yuan2014}. The messages ${\large{\textbf{\emph{e}}}_{\small{EST}}}({\mathbf{x}}_{i})$, ${\tilde {\textbf{\emph{l}}}_{\small{EST}}} ({\mathbf{x}}_{i})$, ${\tilde {\textbf{\emph{l}}}_{\small{DEC}}} ({\mathbf{x}}'_{i})$, and ${\large{\textbf{\emph{e}}}_{\small{DEC}}}({\mathbf{x}}'_{i})$, $i\in\mathcal{N}_u$, are defined as the estimates of the transmissions. As illustrated in Fig. \ref{f2}, at the BS, the received signals $\mathbf{Y}=[\mathbf{y}_1,\cdots,\mathbf{y}_N]$ and message  $\{{\tilde {\textbf{\emph{l}}}_{\small{EST}}} ({\mathbf{x}}_{i}),i\in \mathcal{N}_u\}$ are passed to a LMMSE detector to estimate message ${\large{\textbf{\emph{e}}}_{\small{EST}}}({\mathbf{x}}_{i})$ for decoder $i$, which is then deinterleaved with $\Pi_i^{-1}$ into ${\tilde {\textbf{\emph{l}}}_{\small{DEC}}} ({\mathbf{x}}'_{i})$, $i\in \mathcal{N}_u$. The corresponding single-user decoder outputs message ${\large{\textbf{\emph{e}}}_{\small{DEC}}}({\mathbf{x}}'_{i})$ based on ${\tilde {\textbf{\emph{l}}}_{\small{DEC}}} ({\mathbf{x}}'_{i})$. Similarly, this message is interleaved by $\Pi_i$ to obtain ${\tilde {\textbf{\emph{l}}}_{\small{EST}}} ({\mathbf{x}}_{i})$ for the detector. This process is repeated iteratively until the maximum number of iterations is achieved.

In the rest of this paper, we will not distinguish $\mathbf{x}_i$ and $\mathbf{x}'_i$ as they are same sequences with different permutations, i.e., ${\tilde {\textbf{\emph{l}}}_{\small{DEC}}} ({\mathbf{x}}'_{i})$ and ${\large{\textbf{\emph{e}}}_{\small{DEC}}}({\mathbf{x}}'_{i})$ can be denoted with ${\large{\textbf{\emph{e}}}_{\small{EST}}}({\mathbf{x}}_{i})$ and ${\tilde {\textbf{\emph{l}}}_{\small{EST}}} ({\mathbf{x}}_{i})$. In fact, the messages ${\large{\textbf{\emph{e}}}_{\small{EST}}}({\mathbf{x}}_{i})$ and ${\tilde {\textbf{\emph{l}}}_{\small{EST}}} ({\mathbf{x}}_{i})$  can be replaced by the means and variances respectively.

\subsubsection{Key Assumptions} For simplicity, we make the following assumptions, which are widely used in iterative decoding and turbo equalization algorithms \cite{Richardson2001, Brink2001, Guo2008, Yuan2014}.

\emph{Assumption 1: For the LMMSE detector, each $x_i(t)$ is independently chosen from $\mathcal{S}$ for any $i$ and $t$; the messages $\{{\large{\textbf{\emph{e}}}_{\small{EST}}}({\mathbf{x}}_{i}), i\in {\mathcal{N}}_u\}$ are independent with each other, and the entries of ${\large{\textbf{\emph{e}}}_{\small{EST}}}({\mathbf{x}}_{i})$ are i.i.d. given $\mathbf{x}_i$.}

\emph{Assumption 2: For the decoder, the messages ${\large{\textbf{\emph{e}}}_{\small{DEC}}}({\mathbf{x}}'_{i})$, $i\in {\mathcal{N}}_u$ are independent with each other, and the entries of ${\large{\textbf{\emph{e}}}_{\small{DEC}}}({\mathbf{x}}'_{i})$ are i.i.d. given $\mathbf{x}'_i$.}

Assumptions 1 and 2 decompose the overall process into the local processors such as the detector and decoders, which simplifies the analysis of the iterative process. In detail, Assumption 1 simplifies the LMMSE estimation (see Section II-D-1), and Assumption 2 simplifies the transfer function of decoders (see Section II-C-2).

\subsubsection{\emph{A-posteriori Probability} (APP) Decoder}
We assume each decoder employees APP decoding\footnote{Although computational complexity of the APP decoding is too high to apply in practical systems, low-complexity message-passing algorithms can be used to achieve near-optimal performance \cite{Richardson2001}. APP decoding assumption is included to simplify our analysis. } at the receiver. The  extrinsic variance output of APP decoder is defined as
\begin{equation}\label{e23}
 v_{i,t}= \mathrm{MMSE}\big(x_{i,t}|{\tilde {\textbf{\emph{l}}}_{\small{DEC}}} ({\mathbf{x}}_{i,\sim t})\big).
\end{equation}
From Assumption 2, we have $ v_{i,t}=v_i,\forall t$. Therefore, we can define the \emph{SINR-Variance} transfer function of the decoders as
\begin{equation}\label{e25}
\mathbf{v}_{\bar{\mathbf{x}}}=\bm{\psi}(\bm{\rho}),
\end{equation}
Where $\bm{\psi}(\bm{\rho})=[\psi_1(\rho_1),\cdots,\psi_{N_u}(\rho_{N_u})]$.

\subsection{LMMSE detector}
In the MIMO-NOMA, the complexity of the optimal MAP detector is too high, and LMMSE detector is an alternative low-complexity detector.
\subsubsection{\emph{A-posteriori} LMMSE Estimation}
Message ${\tilde {{\emph{l}}}_{\small{EST}}} ({{x}}_{i,t})$ is de-mapped to ${{\bar {{x}}}_{i,t}}$ with variance ${{v}}_i$. Assumption 1 indicates that ${v}_i$ is invariant with respect to $t$. Hence,
\begin{equation}\label{e7}
\!\!\!\!\!{{\bar x}_{i,t}} \!=\! \mathrm{E}\!\left[{x_{i,t}}|{\tilde {{\emph{l}}}_{\small{EST}}} ({{x}}_{i,t})\right]\!,{v}_i \!=\! \mathrm{E}\!\left[|{x_{i,t}}\!-\!\bar{x}_{i,t}|^2|{\tilde {{\emph{l}}}_{\small{EST}}} ({{x}}_{i,t})\right]\!,
\end{equation}
where $\mathrm{E}[a|b]$ denotes the expectation of $a$ given $b$. Let $\bar{\mathbf{x}}(t)=[{\bar{x}_{1,t}},\cdots,\bar{x}_{N_u,t}]$ and $\mathbf{V}_{\bar{\mathbf{x}}(t)}\!=\!\mathbf{V}_{\bar{\mathbf{x}}}= \mathrm{diag}(v_1,v_2,\cdots,v_{N_u})$. The \emph{a-posteriori} LMMSE estimation \cite{tse2005, Yuan2014, Lei2016, Lei20162} is
\begin{equation}\label{GMP2}
{{\hat {\mathbf{x}}}(t)}
= \mathbf{V}_{\hat {\mathbf{x}}}\left[\mathbf{V}_{\bar{\mathbf{x}}}^{-1}\bar{\mathbf{x}}(t)+ \sigma^{-2}_n\mathbf{H}'^H\mathbf{y}_t  \right],
\end{equation}
where $\mathbf{V} _{{{\hat {\mathbf{x}}}}} = (\sigma _{{{n}}}^{- 2}\mathbf{H}'^H\mathbf{H}'+\mathbf{V} _{{{\bar{\mathbf{x}}}}}^{-1})^{-1}$ denotes the \emph{a-posteriori} deviation of the estimation. A derivation of \eqref{GMP2} is given in APPENDIX \ref{APP:LMMSE}. For more details of LMMSE, please refer to Section II-C-2 and Section IV-F of \cite{Lei2016}.

\subsubsection{Extrinsic LMMSE Detector} Let $\hat{x}_{i,t}$ and $v_{\hat{x}_i}$ be the entry and diagonal entry of ${{\hat {\mathbf{x}}}(t)} $ and $\mathbf{V}_{\hat {\mathbf{x}}}$, respectively. The LMMSE detector outputs extrinsic\footnote{The \emph{a-posteriori}
estimate in \eqref{GMP2} cannot be used directly due to the correlation
issue.} mean and variance for $x_{i,t}$ (denoted by $u_{i,t}$ and $\phi_i^{-1}$) by excluding the prior message ${\tilde {{\emph{l}}}_{\small{EST}}} ({{x}}_{i,t})$ with the message combining rule \cite{Loeliger2006}:
\begin{equation}\label{e9}
\phi_i(\mathbf{v}_{\bar{\mathbf{x}}}) = v_{\hat{x}_i}^{-1}(\mathbf{v}_{\bar{\mathbf{x}}})-{v}_i^{-1}\;\; \mathrm{and} \;\; {u_{i,t}}=\frac{\hat{x}_{i,t}}{{\phi_i}v_{\hat{x}_i}}- \frac{\bar{x}_{i,t}}{{\phi_i}{v}_{i}},
\end{equation}
where $\mathbf{v}_{\bar{\mathbf{x}}}\!=\! [v_1,v_2,\cdots,v_{N_u}]$.

\subsubsection{Extrinsic Transfer Function}
The following proposition is proved in APPENDIX \ref{ASS3_AP}.

\emph{Proposition 1 \cite{Kay1993, Poor1997}: Let $\bm{\rho}=[\rho_1,\cdots,\rho_{N_u}]$, $\bm{\phi}(\mathbf{v}_{\bar{\mathbf{x}}})= \left[{\phi_1}(\mathbf{v}_{\bar{\mathbf{x}}}),\cdots, {\phi_{N_u}}(\mathbf{v}_{\bar{\mathbf{x}}}) \right]$. The output of the LMMSE detector is an observation from AWGN channel\footnote{The "*" indicates that it is not the channel noise, but an imagined noise including the interference.}, i.e., $\mathbf{u}_{t}=\mathbf{x}(t)+\mathbf{n}_t^{*}$ with Signal Interference Noise Ratio (SINR) $\bm{\rho}=\bm{\phi}(\mathbf{v}_{\bar{\mathbf{x}}})$.}

With Proposition 1, we can define the extrinsic LMMSE \emph{SINR-Variance} transfer function of user $i$ as
\begin{equation}
\phi_i(\mathbf{v}_{\bar{\mathbf{x}}})=v_{\hat{x}_i}^{-1}-v_i^{-1},\;\; \mathrm{for}\;\; i\in \mathcal{N}_u.
\end{equation}
The \emph{a-posteriori} MSE of LMMSE detector for user $i$ is
\begin{equation}\label{e19}
\mathrm{mmse}_{ap,i}^{est}(\mathbf{v}_{\bar{\mathbf{x}}})=v_{\hat{x}_i}.
\end{equation}
Furthermore, Proposition 1 will be used to derive the area properties of MIMO-NOMA (see Section III-B).

{\textbf{Remark:}} The variance $v_i$ varies from $0$ to $1$, because the signal power is normalized to 1. From (\ref{e18}), the output estimation of user $i$ depends on the input variances of all the users. Thus, the \emph{SINR-Variance} transfer functions of all users interfere with each other. In addition, $\phi_i (\mathbf{v}_ {\bar{\mathbf{x}}})$ is monotonically decreasing in $\mathbf{v}_{\bar{\mathbf{x}}}$, which means the lower input variances of the users, the higher the output \emph{SINR} of the detector.
\subsection{Complexity of Iterative LMMSE Detection}
From \eqref{GMP2}, the complexity of LMMSE estimator is $\Xi_{est}=\mathcal{O}\left(\,\min\{N_rN_u^2+N_u^3, \;N_u N_r^2+N_r^3\}\right)$, where $\mathcal{O}(N_u^3)$ (or  $\mathcal{O}(N_r^3)$) arises from the matrix inverse calculation, $\mathcal{O}(N_rN_u^2)$ (or $\mathcal{O}(N_uN_r^2)$) from the matrix multiplication, and ``$\min$" from \emph{Matrix Inversion Lemma}. Hence, the total complexity of iterative LMMSE detection is $\mathcal{O}\left((\Xi_{est}+N_u\Xi_{dec})N_{ite}\right)$, where $N_{ite}$ is the number of iterations and $\Xi_{dec}$ denotes the single-user decoding complexity per iteration. Note that the complexity of LMMSE detector is much lower than the optimal MUD whose complexity grows exponentially with $N_u$ and $N_r$, and polynomially with $|\mathcal{S}|$.



\section{Matching Conditions and Area Theorems}
In \cite{Guo2005, Bhattad2007,Yuan2014}, the \emph{I-MMSE} theorem and the area theorems for the P2P communication systems are proposed. In this section, these results are generalized to the MIMO-NOMA systems. 

\subsection{Matching Conditions of MIMO-NOMA}
\subsubsection{SINR-Variance Transfer Chart} The iterative receiver performs iteration between the detector and the decoders, which are described by $\bm{\rho}=\bm{\phi}(\mathbf{v}_{\bar{\mathbf{x}}})$ and  $\mathbf{v}_{\bar{\mathbf{x}}}=\bm{\psi}(\bm{\rho})$ respectively. Hence, the iteration is tracked by
\begin{equation}\label{e27}
\!\!\!\bm{\rho}(\tau)\!=\!\bm{\phi}\left(\mathbf{v}_{\bar{\mathbf{x}}}(\tau-1)\right), \mathbf{v}_{\bar{\mathbf{x}}}(\tau)\!=\!\bm{\psi}\left(\bm{\rho}(\tau)\right), \tau=1,2,\cdots\!.
\end{equation}
Eq. \eqref{e27} converges to a fixed point $\mathbf{v}_{\bar{\mathbf{x}}}^*$, which satisfies\vspace{-0.1cm}
\begin{equation*}\label{e28}
\bm{\phi}\!\left(\mathbf{v}_{\bar{\mathbf{x}}}^*\right) \!=\!\bm{\psi}^{-1}\!\left(\mathbf{v}_{\bar{\mathbf{x}}}^*\right)\; \mathrm{and} \; \bm{\phi}\left(\mathbf{v}_{\bar{\mathbf{x}}}\right) \!>\!\bm{\psi}^{-1}\left(\mathbf{v}_{\bar{\mathbf{x}}}\right),\; \mathrm{for} \; \mathbf{v}_{\bar{\mathbf{x}}}^*  \!<\! \mathbf{v}_{\bar{\mathbf{x}}} \!\leq\! \mathbf{1},\vspace{-0.15cm}
\end{equation*}
where $\bm{\psi}^{-1}(\cdot)$ denotes the inverse of $\bm{\psi}(\cdot)$, which exists since $\bm{\psi}(\cdot)$ is continuous and monotonic \cite{Guo2011}. The inequality\footnote{In this paper, all the inequalities for the vectors or matrixes correspond to the component-wise inequalities.} $\mathbf{v}_{\bar{\mathbf{x}}} \leq \mathbf{1}$ comes from the normalized signal power of $\mathbf{x}(t)$, $t\in \mathcal{N}$.

\begin{figure}[t]\vspace{-0.1cm}
  \centering
  \includegraphics[width=6.5cm]{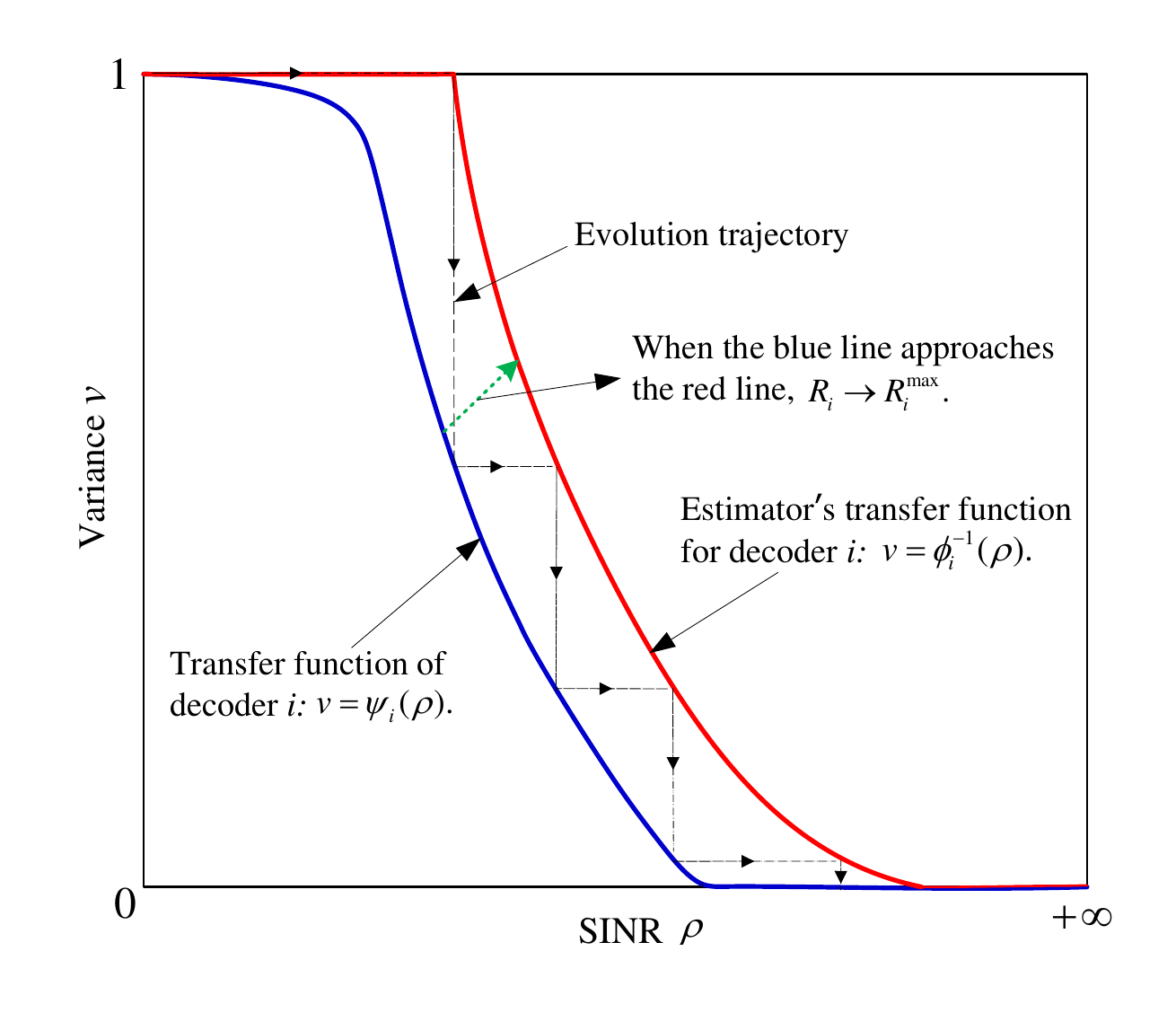}\\\vspace{-0.1cm}
  \caption{SINR-variance transfer chart of the iterative receiver. }\label{area}\vspace{-0.2cm}
\end{figure}

As shown in Fig. \ref{area}, if $\mathbf{v}_{\bar{\mathbf{x}}}^*=\mathbf{0}$, then all the transmissions can be correctly recovered, which means that $\bm{\phi}\left(\mathbf{v}_{\bar{\mathbf{x}}}\right) >\bm{\psi}^{-1}\left(\mathbf{v}_{\bar{\mathbf{x}}}\right)$ for any available $\mathbf{v}_{\bar{\mathbf{x}}}$, i.e., decoders' transfer function $\bm{\psi}^{-1}\left(\mathbf{v}_{\bar{\mathbf{x}}}\right)$ lies below that of the detector $\bm{\phi}\left(\mathbf{v}_{\bar{\mathbf{x}}}\right)$.

\subsubsection{Matching Conditions} The detector and decoders are matched if\vspace{-0.15cm}
\begin{equation}\label{e29}
\bm{\phi}\left(\mathbf{v}_{\bar{\mathbf{x}}}\right) =\bm{\psi}^{-1}\left(\mathbf{v}_{\bar{\mathbf{x}}}\right),\quad \mathrm{for} \;\; \mathbf{0}  < \mathbf{v}_{\bar{\mathbf{x}}} \leq \mathbf{1}.\vspace{-0.15cm}
\end{equation}

Therefore, we obtain the following proposition.

\emph{Proposition 2}: \emph{For any $i\in \mathcal{N}_u$, the matching conditions of the iterative MIMO-NOMA systems can be rewritten to}
\begin{eqnarray}
\psi_i(\rho_i)&=&\phi_i^{-1}(\phi_i(\mathbf{1}))=1, \;\;\mathrm{for}\;\; 0\leq\rho_i<\phi_i(\mathbf{1});\label{e30}\\
\psi_i(\rho_i)&=&\phi_i^{-1}(\rho_i),\;\;\mathrm{for}\;\; \phi_i(\mathbf{1})\leq\rho_i<\phi_i(\mathbf{0});\label{e31}\\
\psi_i(\rho_i)&=&0,\;\;\mathrm{for}\;\; \phi_i(\mathbf{0})\leq\rho_i<\infty.\label{e32}
\end{eqnarray}
\begin{IEEEproof}
Eq. \ref{e29} means that $\phi_i(\mathbf{v}_{\bar{\mathbf{x}}})=\psi_i^{-1}(v_i)$ for any $i\in \mathcal{N}_u$. First, we have $\phi_i(\mathbf{1})>0$, since the detector always uses the information from the channel. Hence, we get $\psi_i(\rho_i)=1$, for $ 0\leq\rho_i<\phi_i(\mathbf{1})$. Second, we have $\phi_i(\mathbf{0})>1$, since the detector cannot remove the uncertainty introduced by the channel noise. Hence, we get $\psi_i(\rho_i)=0,\;\;\mathrm{for}\;\; \phi_i(\mathbf{0})\leq\rho_i<\infty$. At last, $\psi_i(\rho_i)=\phi_i^{-1}(\rho_i)$ exists due to its monotonicity on $\phi_i(\mathbf{1})\leq\rho_i<\phi_i(\mathbf{0})$. Therefore, we have \eqref{e30}-\eqref{e32}.
\end{IEEEproof}
Proposition 2 will be used in the area properties and rate analysis of MIMO-NOMA.

\subsection{Area Properties}
Let $\mathrm{snr}_{pri,i}^{dec}$ denote the \emph{SNR} of the a-priori message for decoder $i$, $\mathrm{snr}_{ext,i}^{est}$ be the \emph{SNR} of the extrinsic message for user $i$ at detector, $\mathrm{mmse}_{ap,i}^{est}(\cdot)$ be the \emph{a-posteriori} variance of the message for user $i$ at detector, and $\mathrm{mmse}_{ap,i}^{dec}(\cdot)$ be the \emph{a-posteriori} variance of the message at decoder $i$.
Besides, $\mathbf{snr}_{ext,i}^{est}=[\mathrm{snr}_{ext,1}^{est}, \cdots,\mathrm{snr}_{ext,N_u}^{est}]$. The following proposition gives the area properties of the iterative detection, which will be used to derive the user rate of MIMO-NOMA.

 \emph{Proposition 3}: \emph{The achievable rate $R_i$ of user $i$ and an upper bound of $R_i$ are given by}\vspace{-0.2cm}
 \begin{eqnarray}
 &&{R_i}  =\int\limits_0^\infty  {\mathrm{mmse}_{ap,i}^{\!dec}({{snr}}_{pri,i}^{dec})d} \mathrm{snr}_{pri,i}^{dec},\\
&&R_i^{\max }=\int\limits_0^\infty {\mathrm{mmse}_{ap,i}^{est}(\mathbf{snr}_{ext}^{est})d} \mathrm{snr}_{ext,i}^{est},\vspace{-0.2cm}
\end{eqnarray}
\emph{where $R_i\leq R_i^{\mathrm{max}}$, $i\in \mathcal{N}_u$, where the equality holds if and only if the \emph{SINR-Variance} transfer functions of the detector and user decoders are matched with each other, i.e., the matching conditions (\ref{e29})$\sim$ (\ref{e32}) hold.}

From (\ref{e18}), (\ref{e25}) and Proposition 1, we have ${\mathrm{snr}_{pri,i}^{dec}}=\rho_i$, $\mathrm{snr}_{ext,i}^{est}=\phi_i(\mathbf{v}_{\bar{\mathbf{x}}})$, $\mathrm{mmse}_{ap,i}^{dec}({{{snr}}_{pri}^{dec,i}})={{\big( {{\rho _i} + {\psi _i}{{({\rho _i})}^{- 1}}} \big)}^{- 1}}$ and $\mathrm{mmse}_{ap,i}^{est}(\mathbf{snr}_{ext,i}^{est})=v_{\hat{x}_i}(\mathbf{v}_{\bar{\mathbf{x}}})$.
Therefore, we have the following corollary from Proposition 3.

\emph{Corollary 1}:\emph{With the \emph{SINR-Variance} transfer functions $\bm{\rho}=\bm{\phi}(\mathbf{v}_{\bar{\mathbf{x}}})$ and $\mathbf{v}_{\bar{\mathbf{x}}}=\bm{\psi}(\bm{\rho})$,} the achievable rate $R_i$ of user $i$ and an upper bound of $R_i$ are
 \begin{eqnarray}
{R_i} = \int\limits_0^\infty  {{{\big( {{\rho _i} + {\psi _i}{{({\rho _i})}^{- 1}}} \big)}^{- 1}}d{\rho _i}} ,\label{e35}\\
R_i^{\max } = \int\limits_0^\infty  v_{\hat{x}_i}(\mathbf{v}_{\bar{\mathbf{x}}}) d \phi_i(\mathbf{v}_{\bar{\mathbf{x}}}),\quad\;\;\;\label{e36}
\end{eqnarray}
\emph{respectively, and $R_i\leq R_i^{\mathrm{max}}$, $i\in \mathcal{N}_u$, where the equality holds if and only if the matching conditions (\ref{e29})$\sim$ (\ref{e32}) hold.}

Now, the achievable rates can be calculated by (\ref{e36}) or (\ref{e35}) together with (\ref{e29}) and the matching conditions (\ref{e30})$\sim$ (\ref{e32}). 

\section{Achievable Rate of Iterative LMMSE Detector}
User achievable rate is derived for the iterative MIMO-NOMA in this section. The Superposition Coded Modulation (SCM) code is employed for the Forward Error Correction (FEC) code. We show that the achievable rate of iterative LMMSE can achieve the capacity of symmetric MIMO-NOMA and sum capacity of asymmetric MIMO-NOMA.

\subsection{Achieving the Sum Capacity of Asymmetric MIMO-NOMA}
For a general asymmetric MIMO-NOMA, achievable rate analysis becomes more complicated due to challenges below.
\begin{itemize}
    \item All the users' transfer functions interfere with each other at the detector, i.e., the any output of the detector relies on every variance of the input messages from the decoders.
    \item All the transfer curves of decoders requires to lie below that of the detector.
    \item  The detector and decoders are associated with each other. It is intractable to optimize over an abstract class of decoder transfer functions for each user.
\end{itemize}

\subsubsection{Transfer-Constraint Parameter} The area theorem tells us that the achievable rate of every user is maximized if and only if its transfer function matches with that of the detector. Therefore, we can fix the transfer functions of the detector, and then obtain users' achievable rate by matching the decoders' transfer functions with the detector.

To make the analysis feasible, we consider a \emph{transfer constraint} for the input variances of the detector.
\begin{equation}\label{e45}
\gamma_i (v_i^{-1}-1)=\gamma_j (v_j^{-1}-1), \;\; \mathrm{for \;\; any} \;\;i,j\in \mathcal{N}_u.
\end{equation}
Let $\bm{\gamma} = [\gamma_1, \cdots, \gamma_{N_u}]$ be the \emph{transfer-constraint parameter} of the iterative LMMSE detection. Without loss of generality, we assume $\gamma_1=1$ and $\gamma_i>0$ , that is, $v_i^{-1}=1+\gamma_i^{-1}(v_1^{-1}-1)$ for any $i\in \mathcal{N}_u$.

Actually, different values of $\bm{\gamma}$ give different variance tracks. Furthermore, different variance tracks correspond to different achievable rates of the users, i.e., the user's achievable rate can be adjusted by the \emph{transfer-constraint parameter} $\bm{\gamma}$.

\begin{figure}[t]\vspace{-0.3cm}
  \centering
  \includegraphics[width=7.0cm]{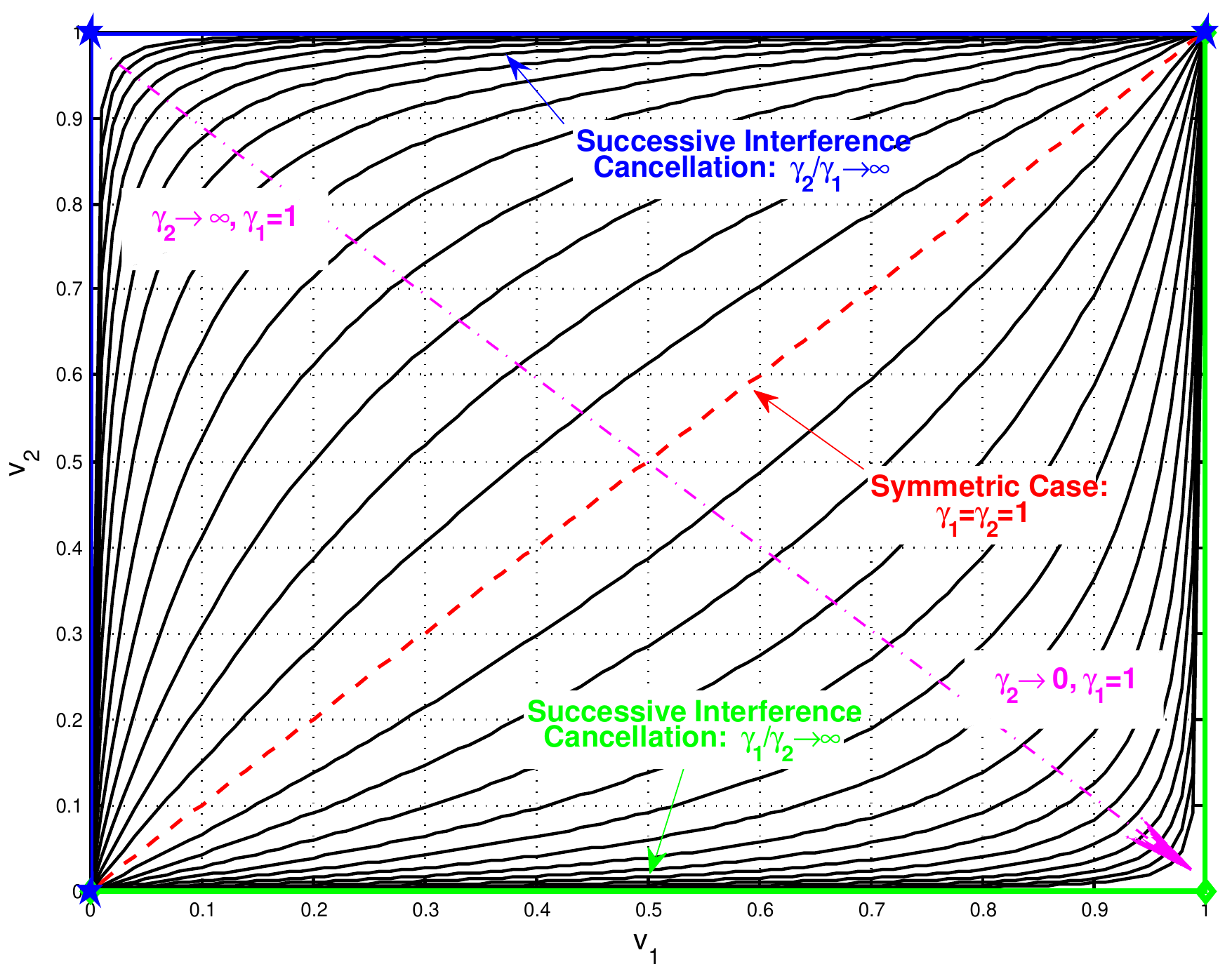}\\\vspace{-0.2cm}
  \caption{Variance tracks for different $\bm{\gamma}$, where $\gamma_1=1$ is fixed. $v_i$ denotes the variance of user $i$, $i=1,2$. When $\gamma_2$ changes from $\infty$ to $0$, the track changes from the blue curve (SIC case with decoding order: user $1\to$ user $2$) to green curve (SIC case with decoding order: user $2\to$ user $1$). When $\gamma_1=\gamma_2=1$, it degenerates into the symmetric case (red line). }\label{f3}\vspace{-0.0cm}
\end{figure}

Fig. \ref{f3} and Fig. \ref{f4} presents the variance tracks with different values of $\bm{\gamma}$ for two-users and three-user MIMO-NOMA systems respectively. As we can see, (\ref{e45}) includes the symmetric case (i.e. $w_1=\cdots =w_{N_u}$) and all the SIC points (maximal extreme points of the capacity region). If $\gamma_{k_i}/\gamma_{k_{i-1}}\to \infty$, for any $i\in \mathcal{N}_u/\{1\}$, we obtain the SIC points with the decoding order $[k_1,k_2,\cdots,k_{N_u}]$, which is a permutation of $[1,2,\cdots,N_u]$. The blue curve and green curves in Fig. \ref{f3} and Fig. \ref{f4} correspond to the SIC cases.

\subsubsection{Transfer Function}
With the \emph{transfer constraint} in (\ref{e45}), we have
\begin{equation}\label{e46}
\mathbf{V}_{\bar{\mathbf{x}}}^{-1}= \mathbf{I}_{N_u} + \gamma_i(v_i^{-1}-1)\bm{\Lambda}_{\bm{\gamma}}^{-1} = \mathbf{V}_{\bar{\mathbf{x}}}^{-1}(v_i)
\end{equation}
and\vspace{-0.3cm}
\begin{eqnarray}\label{evall}
\mathbf{V}_{\hat{\mathbf{x}}}&=&(\sigma _{{{n}}}^{- 2}\mathbf{H}'^H\mathbf{H}'+\mathbf{V} _{{{ \bar{\mathbf{x}}}}}^{-1})^{-1}\nonumber\\
&=& (\sigma _{{{n}}}^{- 2}\mathbf{H}'^H\mathbf{H}'+ \mathbf{V}_{\bar{\mathbf{x}}}^{-1}(v_i))^{-1}\nonumber\\
&=& \mathbf{V}_{\hat{\mathbf{x}}}(v_i)
\end{eqnarray}
where $i\in\mathcal{N}_u$, and $\bm{\Lambda}_{\bm{\gamma}}=\mathrm{diag}(\bm{\gamma})$ is a diagonal matrix whose diagonal entries are $\bm{\gamma}$. Thus, we have
\begin{equation}\label{ephi}
\phi_i(\mathbf{v}_{\bar{\mathbf{x}}})= v_{\hat{x}_i}(v_i)^{-1}-v_i^{-1}=\phi_i(v_i)=\rho_i.
\end{equation}
For example, if we take $i=1$, we have
\begin{equation}\label{ev_1}
\mathbf{V}_{\bar{\mathbf{x}}}^{-1}\!=\! \mathbf{V}_{\bar{\mathbf{x}}}^{-1}(v_1),\;
\mathbf{V}_{\hat{\mathbf{x}}} \!=\!  \mathbf{V}_{\hat{\mathbf{x}}}(v_1),\;\mathrm{and}\;\;
\phi_i(\mathbf{v}_{\bar{\mathbf{x}}})\!=\! \phi_i(v_1).
\end{equation}
\subsubsection{Asymmetric Matching Condition} With the \emph{transfer constraint}, the matching conditions are simplified as follows.

\emph{Proposition 5: Based on (\ref{ephi}), \emph{for $i\in \mathcal{N}_u$, the matching conditions (\ref{e29}) can be rewritten to}
\begin{eqnarray}
\psi_i(\rho_i)\!\!\!\!&=&\!\!\!\!\phi_i^{-1}(\phi_i({1}))=1, \;\;\mathrm{for}\;\; 0\leq\rho_i<\phi_i({1});\label{em1}\\
\psi_i(\rho_i)\!\!\!\!&=&\!\!\!\!\phi_i^{-1}(\rho_i),\;\;\mathrm{for}\;\; \phi_i({1})\leq\rho_i<\phi_i({0});\label{em2}\\
\psi_i(\rho_i)\!\!\!\!&=&\!\!\!\!0,\;\;\mathrm{for}\;\; \phi_i({0})\leq\rho_i<\infty.\label{em3}
\end{eqnarray}}\vspace{-0.3cm}
\begin{IEEEproof}
{From \eqref{ev_1}, we have $\phi_i({\mathbf{1}})=\phi_i({1})$ and $\phi_i({\mathbf{0}})=\phi_i({0})$. Substituting it to \eqref{e30}-\eqref{e32}, we obtain Proposition 5.}
\end{IEEEproof}
\subsubsection{User Achievable Rate}
The users' achievable rates are given by the following lemma.
\begin{figure}[t]
  \centering
  \includegraphics[width=9.5cm]{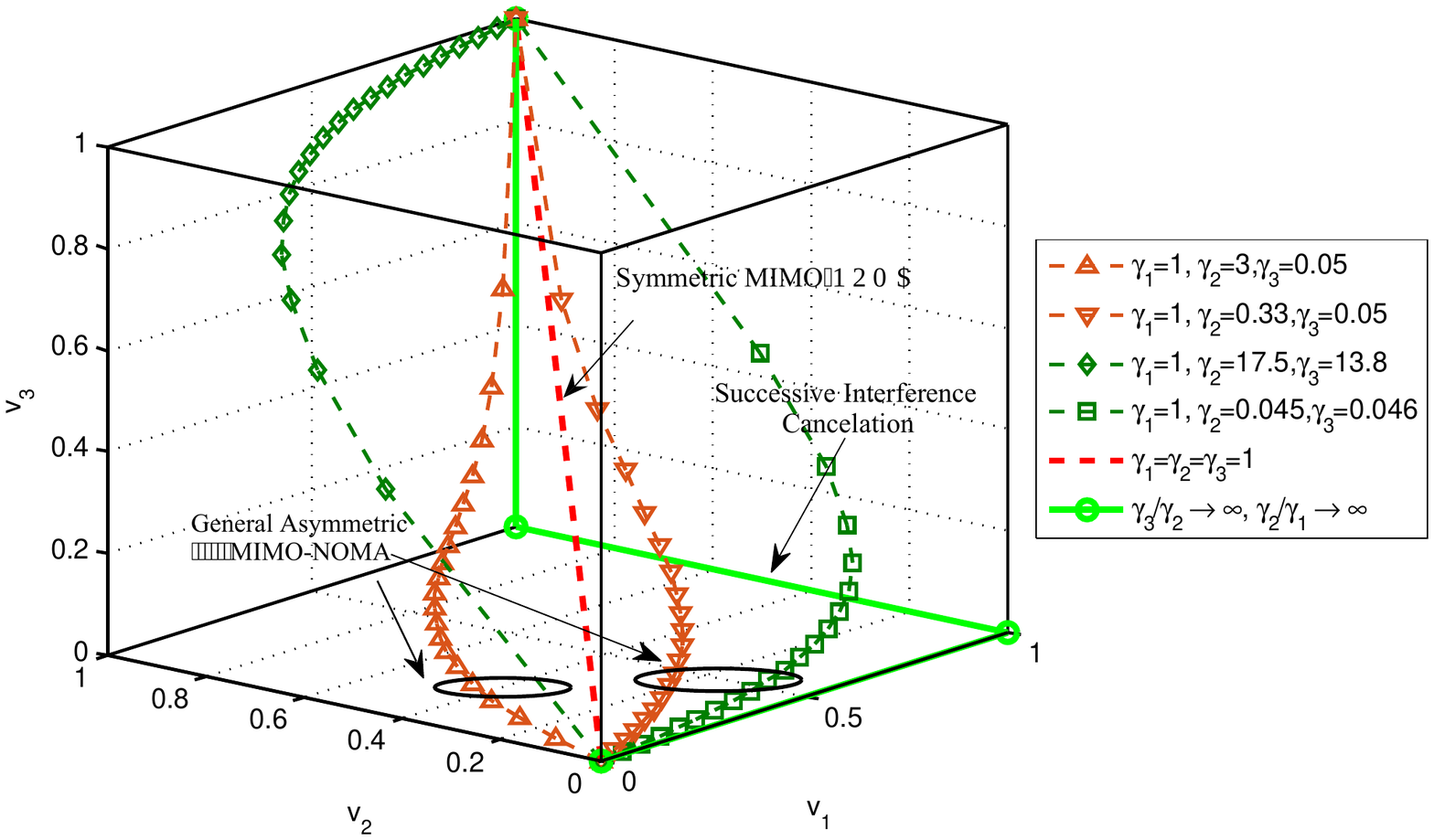}\\\vspace{-0.1cm}
  \caption{Variance tracks for different $\bm{\gamma}$, where $\gamma_1=1$ is fixed. $v_i$ denotes the variance of user $i$, $i=1,2,3$. The variance track changes with $\gamma_2$ and $\gamma_3$. When $\gamma_3/\gamma_2\to\infty$ and $\gamma_2/\gamma_1\to\infty$ (green curve),  it degenerates into the SIC case with the decoding order: user $3\to$user $2\to$ user $1$. When $\gamma_1=\gamma_2=\gamma_3=1$, it degenerates into the symmetric case (red line). The other curves are the general asymmetric cases.}\label{f4}\vspace{-0.3cm}
\end{figure}

\emph{\textbf{Lemma 1}: For the asymmetric MIMO-NOMA {with any $N_u$ and $N_r$}, the achievable rate of user $i$ for iterative LMMSE detection is
\begin{equation}\label{lemma1}
R_i = \int\limits_{v_1=1}^{v_1=0} \left[v_1- \gamma_i^{-1} \left[\mathbf{V}_{\hat{\mathbf{x}}}(v_1)\right]_{i,i} \right] dv_1^{-1}
- \log(\gamma_i),
\end{equation}
where $\mathbf{V}_{\hat{\mathbf{x}}}(v_1)=\left( \sigma^{-2}_n\mathbf{H}'^H\mathbf{H}'+ \mathbf{I}_{N_u} + (v_1^{-1}-1)\bm{\Lambda}_{\bm{\gamma}}^{-1}\right)^{-1}$, and $[\cdot]_{i,i}$ denotes the $i$-th diagonal entry of the matrix.}

\begin{IEEEproof}
See APPENDIX \ref{APP_Lemma1}.
\end{IEEEproof}

{Lemma 1} gives the achievable rate of each user, but it is an complicated integral function and we cannot see the specific relationship between the achievable rates and $\Lambda_{\bm{\gamma}}$.

{\emph{Remark:}} When $\gamma_i=1$ for $i\in \mathcal{N}_u$, and for a symmetric system with: \emph{(i)} the same rate $R_i=R$ for $i\in \mathcal{N}_u$; \emph{(ii)} the same power $\mathbf{K}_{\mathbf{x}}\!=\!w^2\mathbf{I}$, \emph{Theorem 1} degenerates to \emph{Corollary 2}.

\subsubsection{Achievable Sum Rate} Although it is difficult to give the exact achievable rate region, the iterative LMMSE detection is shown to sum capacity achieving.

\emph{\textbf{Theorem 1:} {For any $N_u$ and $N_r$}, the iterative LMMSE detection achieves the sum capacity of MIMO-NOMA, i.e., $R_{sum}=\log |\textbf{I}_{N_u} + {\sigma_n^{-2}}\mathbf{H}'\mathbf{H}'^H|$.}

\begin{IEEEproof}
See APPENDIX \ref{APP_Theorem2}.
\end{IEEEproof}

Theorem 1 shows that for a general asymmetric MIMO-NOMA, from the sum rate perspective, the LMMSE detector is an optimal detector without losing any useful information during the estimation.

\subsubsection{Monotonicity of Achievable Rate} The following lemma shows the monotonicity of achievable rate in \eqref{lemma1}.

\emph{\textbf{Lemma 2}: The achievable rate $R_i$ of user $i$ increases monotonously with $\gamma_i$ and decreases monotonously with $\gamma_j$, where $i,j \in \mathcal{N}_u$ and $j\neq i$.}
\begin{IEEEproof}
It is easy to find that $\mathrm{mmse}^{est}_{ap,i}$ (or $\mathrm{mmse}^{dec}_{ap,i}$) increases monotonously with $\gamma_i$ and decreases monotonously $\gamma_j$ for $i,j \in \mathcal{N}_u$ and $j\neq i$. Thus, based on \emph{Proposition 3}, we have that $R_i$ increases monotonously with $\gamma_i$ and decreases monotonously $\gamma_j$ for $j\neq i$.
\end{IEEEproof}

Lemma 2 is important in user rate adjustment, i.e., if we want increase the rate of user $i$, it only needs to increase $\gamma_i$. Besides, the monotonicity is also important for the practical iterative detection design.

\subsection{Achieving the Capacity of Symmetric MIMO-NOMA}
Then, we consider a simple symmetric MIMO-NOMA systems, that is the users have the same power and the same rate, i.e., $\mathbf{K}_{\mathbf{x}}=w^2\mathbf{I}$ and $R_i=R_j$, for $ i,j\in\mathcal{N}_u$.

\subsubsection{Transfer Function} Since all the users have the same conditions, we thus obtain that all the users have the same transfer functions, which means $v_i=v$ and $\rho_i=\rho$, for any $i\in\mathcal{N}_u$. Therefore, the transfer functions are derived as: \vspace{-0.0cm}
 \begin{eqnarray}\label{e37}
\!\!\!\!\!\!v_{\hat{x}_i}(\mathbf{v}_{\bar{\mathbf{x}}}) \!\!\!\!&\mathop  =\limits^{(a)} & \!\!\!\! \frac{1}{N_u}\mathrm{mmse}_{ap}^{est}(\mathbf{v}_{\bar{\mathbf{x}}})= \frac{1}{N_u} \mathrm{Tr}\{\mathbf{V} _{{{\hat {\mathbf{x}}}}}\} \nonumber\\
&=&\!\!\!\!\frac{1}{N_u} \mathrm{Tr}\{\left(\sigma _{{{n}}}^{- 2}w^2\mathbf{H}^H\mathbf{H}+v^{-1}\mathbf{I}_{N_u}\right)^{-1}\}\nonumber\\
&=&\!\!\!\!v_{\hat{x}}(v),
\end{eqnarray}\vspace{-0.1cm}
and\vspace{-0.1cm}
\begin{eqnarray}\label{e38}
\!\!\!\!\!\!\phi_i(\mathbf{v}_{\bar{\mathbf{x}}})
&\!\!\!\!\!\!\mathop  =\limits^{(b)}\!\!\!\!\!\! & \!\!\! {v_{\hat{x}}(v)}^{-1}-v^{-1} \nonumber\\
&=& \!\!\!\!\frac{1}{N_u} \mathrm{Tr}\{\left(\sigma _{{{n}}}^{- 2}w^2\mathbf{H}^H\mathbf{H}+v^{-1}\mathbf{I}_{N_u}\right)^{-1}\}^{-1} \!-\! v^{-1}\nonumber\\
&=&\!\!\!\phi(v)=\rho,
\end{eqnarray}
where equations (a) and (b) are obtained from the symmetric assumption. Similarly, we have $\psi_i(\rho_i)=\psi(\rho)$, $i\in \mathcal{N}_u$.

\subsubsection{Matching Condition} Since all the users are symmetric, Proposition 2 can be simplified as follows.

\emph{Proposition 4}: \emph{The matching conditions of the iterative symmetric MIMO-NOMA system are given by}\vspace{-0.2cm}
\begin{eqnarray}
\psi(\rho)&=&\phi^{-1}(\phi({1}))=1, \;\;\mathrm{for}\;\; 0\leq\rho<\phi({1});\label{e39}\\
\psi(\rho)&=&\phi^{-1}(\rho),\;\;\mathrm{for}\;\; \phi({1})\leq\rho<\phi({0});\label{e40}\\
\psi(\rho)&=&0\;\;\mathrm{for},\;\; \phi({0})\leq\rho<\infty.\label{e41}
\end{eqnarray}

\subsubsection{Achievable Rate} In this case, the analysis of symmetric MIMO-NOMA degenerates into that of single-user and single-antenna system. From the transfer functions and matching conditions above, we obtain the following theorem.

\textbf{\emph{Corollary 2}}: \emph{For a symmetric MIMO-NOMA {with any $N_u$ and $N_r$} that: (i) $R_i=R, \forall i\in \mathcal{N}_u$; (ii) $\mathbf{K}_{\mathbf{x}}=w^2\mathbf{I}$; the iterative LMMSE detection achieves the capacity, i.e., $R_{i}=\frac{1}{N_{u}}\log |I_{N_r} + \frac{w^2}{\sigma_n^2}\mathbf{H}\mathbf{H}^H|,\forall i\in \mathcal{N}_u$, and $R_{sum}=\log |I_{N_r} + \frac{w^2}{\sigma_n^2}\mathbf{H}\mathbf{H}^H|$.}


Corollary 2 shows that for a symmetric MIMO-NOMA system, the iterative detection structure is optimal, i.e., the LMMSE detector is an optimal detector without losing any useful information during the estimation.

\subsection{Practical Iterative LMMSE Detection Design}
It should be noted that the codes design depends also on $\Lambda_{\bm{\gamma}}$. Since we cannot get a closed-form solution of the user rate with respect to $\Lambda_{\bm{\gamma}}$, it is hard to obtain the proper $\Lambda_{\bm{\gamma}}$ for the given user rates. Nevertheless, Algorithm 1 provides a numeric solution of $\Lambda_{\bm{\gamma}}$ to satisfy user rate requirement.

\begin{algorithm}
\caption{Algorithm for finding $\Lambda_{\bm{\gamma}}$}
\begin{algorithmic}[1]
\State {\small{\textbf{Input:} {{$\mathbf{H}$,  $\mathbf{K}_{\mathbf{x}}$, $\sigma^2_n$, $\epsilon>0$, $\delta>0$, $N_{max}$}}, $\mathbf{R}=[R_1,\cdots,R_{N_u}]$ and calculate {{$\mathbf{H}'$}}.
\State \textbf{If} $\mathbf{R}\in \mathbf{\mathcal{R}}_\mathcal{S}$ ($\mathbf{\mathcal{R}}_\mathcal{S}$ is the capacity region given by (\ref{e17}))
\State \quad\;\; Random choose $\bm{\gamma}=[\gamma_1,\cdots,\gamma_{N_u}]$, $\gamma_i>0$, $\forall i\in \mathcal{N}_u$,
\Statex \quad\;\; Calculate $\mathbf{R}^{(0)}(\bm{\gamma})=[R_1^{(0)},\cdots,R_{N_u}^{(0)}]$ by (\ref{lemma1}) and $t=1$.
\State \quad\;\; \textbf{While} \;{\small{$\left( \:||\mathbf{R}^{(0)}-\mathbf{R}||_1>\epsilon \;{\textbf{or}}\; t<N_{max} \;\right)$}}
\State \quad\quad\;\; \textbf{For} $i=1:N_u$
\State \quad\quad\quad\;\; fixed $\bm{\gamma}_{\sim i}=[\gamma_1,\cdots,\gamma_{i-1},\gamma_{i-1}, \cdots,\gamma_{N_u}]$, \Statex \quad\quad\quad\;\; find $\gamma_i^*$ for ${R}_i^{(1)}(\gamma_i=\gamma_i^*)={R}_i$, and
\State \quad\quad\quad\;\; calculate $\mathbf{R}^{(1)}(\bm{\gamma}_{\sim i},\gamma_i^*)=[R_1^{(1)},\cdots,R_{N_u}^{(1)}]$.
\State \quad\quad\quad\;\; \textbf{While} $ ||\mathbf{R}^{(1)}-\mathbf{R}||_1>||\mathbf{R}^{(0)}-\mathbf{R}||_1$ \State \quad\quad\quad\quad\quad\;\;$\gamma_i^*=(\gamma_i+\gamma_i^*)/2$ and go to step 7.
\State \quad\quad\quad\;\; \textbf{End While}
\State \quad\quad\quad\;\; $\gamma_i=\gamma_i^*$ and $\mathbf{R}^{(0)}=\mathbf{R}^{(1)}$.
\State \quad\quad\;\; \textbf{End For}
\State \quad\quad\;\; $t=t+1$.
\State \quad\;\; \textbf{End While}
\State \quad\;\; \textbf{If} $t<N_{max}$
\State \quad\quad\;\; \textbf{Output:}  $\bm{\gamma}$.
\State \quad\;\; \textbf{Else}
\State \quad\quad\;\; {$R_i=R_i-\delta$, $\forall i\in \mathcal{N}_u$.}
\State \quad\;\; \textbf{End If}
\State \textbf{Else} $\mathbf{R}\notin \mathbf{\mathcal{R}}_\mathcal{S}$
\State \quad\;\;  Find the projection $\mathbf{R}^*$ of $\mathbf{{R}}$ on the dominant face of $\mathbf{\mathcal{R}}_\mathcal{S}$.
\State \quad\;\; $\mathbf{R}=\mathbf{R}^*$, and go back to step 2.
\State \textbf{End If} }}
\end{algorithmic}
\end{algorithm}

{For any $N_u$ and $N_r$}, Algorithm 1 gives a numeric search of $\Lambda_{\bm{\gamma}}$ given rate $\mathbf{R}$, where $N_{max}$ is the maximum iterative number, $\epsilon$ and $\delta$ indicate the allowed precision, and $||\cdot||_1$ denotes the 1-norm. It should be noted that $\gamma^*_i$ in step 6 definitely exists and can be easy searched by dichotomy or quadratic interpolation method as $R_i$ increases monotonously with $\gamma_i$ (Lemma 2). In addition, steps $8\sim 10$ ensure that the new $\gamma_i^*$ is always better than the previous one and the search program will not stop until the requirement $\Lambda_{\bm{\gamma}}$ is got. Experimentally, we find that the points in the system capacity region are always achievable.

\section{Important Properties and Special Cases of Iterative LMMSE detection}
Can the iterative LMMSE detection achieve all points in the capacity region of asymmetric MIMO-NOMA? To answer this question, we derive some properties and show that:
\begin{itemize}
\item for the two-user MIMO-NOMA, all points in the capacity region can be achieved by iterative LMMSE detection;
\item all the maximal extreme points in the capacity region of MIMO-NOMA {with any number of users} can be achieved by iterative LMMSE detection.
\end{itemize}
Furthermore, MISO and SIMO are discussed as two special cases, which show that the ESE in IDMA and MRC are sum capacity optimal for MISO and SIMO respectively. \vspace{-0.3cm}

\subsection{Achieving the Maximal Extreme Point}
As it is mentioned in \emph{Capacity Region Domination Lemma} in APPENDIX \ref{CR_domination_lemma}, the system capacity region is dominated by a convex combination of the maximal extreme points, which can be achieved by SIC.

Here, we show that all these maximal extreme points can be achieved by iterative LMMSE detection when the transfer-constraint parameter $\Lambda_{\bm{\gamma}}$ is properly chosen.

\textbf{\emph{Corollary 3}}: \emph{{For any $N_u$ and $N_r$}, all the maximal extreme points in the capacity region of MIMO-NOMA can be achieved by iterative LMMSE detection.}

\begin{IEEEproof}
See APPENDIX \ref{APP_Coro1}.
\end{IEEEproof}

This corollary shows that if the parameter $\Lambda_{\bm{\gamma}}$ is properly chosen, the iterative LMMSE detection degenerates into the SIC methods, i.e., the SIC methods are special cases of the proposed iterative LMMSE detection.\vspace{-0.2cm}
\begin{figure}[t]
  \centering
  \includegraphics[width=8.0cm]{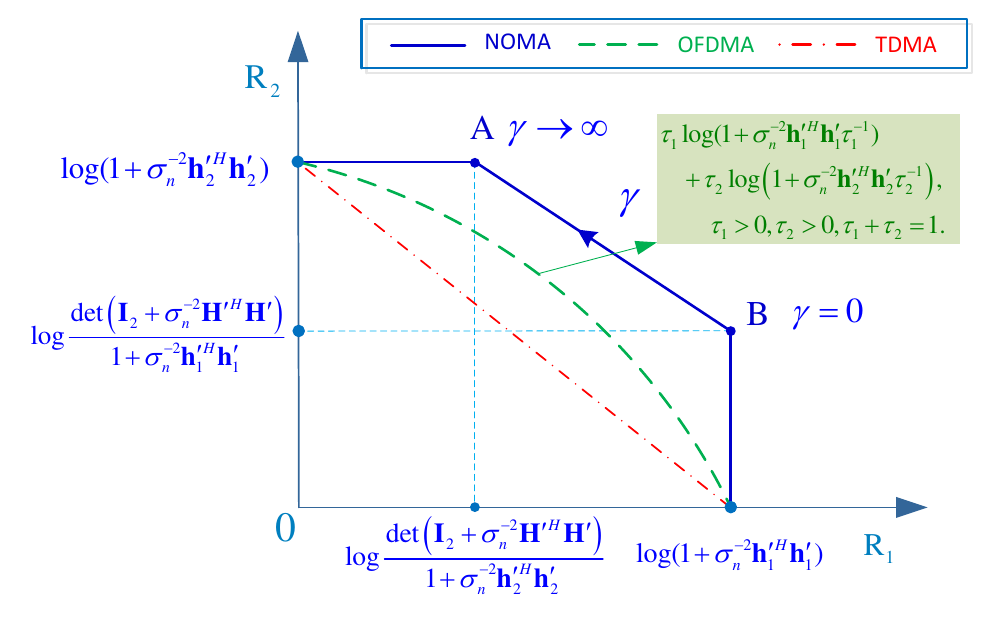}\\\vspace{-0.2cm}
  \caption{Achievable region of iterative LMMSE detection for two-user MIMO-NOMA system. When the parameter $\gamma$ changes from $0$ to $\infty$, point $(R_1,R_2)$ moves from maximal extreme point B to maximal extreme point A along segment AB. }\label{f5}
\end{figure}

\subsection{Two-user MIMO-NOMA}
As it is mentioned, it is hard to calculate the specific achievable user rates from (\ref{lemma1}). However, in two-user case, the achievable rate region can be calculated and it equals to the capacity of MIMO-NOMA.

\emph{\textbf{Theorem 3}}: \emph{Iterative LMMSE detection achieves the whole capacity region of two-user MIMO-NOMA:}
\begin{equation}
\left\{\begin{array}{l}
R_1\leq\log ( {1 + \frac{1}{{\sigma _n^2}}{\mathbf{h}'}_1^H{{\mathbf{h}'}_1}} ),\\
R_2\leq\log ( {1 + \frac{1}{{\sigma _n^2}}{\mathbf{h}'}_2^H{{\mathbf{h}'}_2}} ),\\
R_1+R_2\leq\log | \mathbf{I}_{2} + {\sigma _n^{- 2}}\mathbf{H}'^H \mathbf{H}'|.
\end{array} \right.
\end{equation}

\begin{IEEEproof}
The pentagon in Fig. \ref{f5} indicates the capacity region of two-user MIMO-NOMA system, which is dominated by segment AB, and point A and point B are two maximal extreme points. Without loss of generality, we let $\gamma_1=1$ and $\gamma_2=\gamma\in[0,\infty)$. From \emph{Theorem 1}, we get
\begin{equation}\label{twouser1}
R_{sum}=R_1+R_2=\log |\mathbf{I}_{2} + {\sigma _n^{- 2}}\mathbf{H}'^H \mathbf{H}'|,
\end{equation}
which is the exact sum capacity of the system.

In addition, as we discussed in \emph{Corollary 3}, when $\gamma$ changes from $0$ to $\infty$,  $R_1$ reduces from $\log ( {1 + \frac{1}{{\sigma _n^2}}{\mathbf{h}'}_1^H{{\mathbf{h}'}_1}} )$ to $\log | \mathbf{I}_2 + {\sigma _n^{- 2}}\mathbf{H}'^H\mathbf{H}'| -\log ( {1 + \frac{1}{{\sigma _n^2}}{\mathbf{h}'}_1^H{{\mathbf{h}'}_1}} )$, and $R_2$ increases from $\log | \mathbf{I}_2 + {\sigma _n^{- 2}}\mathbf{H}'^H\mathbf{H}' |-\log ( {1 + \frac{1}{{\sigma _n^2}}{\mathbf{h}'}_2^H{{\mathbf{h}'}_2}} )$ to $\log ( {1 + \frac{1}{{\sigma _n^2}}{\mathbf{h}'}_2^H{{\mathbf{h}'}_2}} )$. As the $R_1$ and $R_2$ are both continuous functions of $\gamma$, from (\ref{twouser1}), we can see that when the parameter $\gamma$ changes from $0$ to $\infty$, the point $(R_1,R_2)$ moves from maximal extreme point B to maximal extreme point A along the segment AB. It means that the iterative LMMSE detection can achieve any point on the segment AB. Therefore, the iterative LMMSE detection achieves all points in the capacity region as it is dominated by the segment AB.
\end{IEEEproof}

Let $\gamma_1=1$ and $\gamma_2=\gamma$, and we can give the specific expressions of $R_1$ and $R_2$. The following corollary is derived directly from \emph{Lemma 1}.

\emph{\textbf{Corollary 4}: For two-user MIMO-NOMA with iterative LMMSE detection, the user rates are given by}\vspace{-0.1cm}
\begin{equation}\label{coro2_1}
\!\!\!\!\!\left\{\!\!\!\begin{array}{l}
R_1=\frac{1}{2}\log(\gamma|A|) +\frac{a_{22}\gamma-a_{11}}{2\eta} \log\frac{a_{22}\gamma+a_{11}-\eta}{a_{22}\gamma+a_{11}+\eta},\\
R_2= \frac{1}{2}\log(\gamma^{-1}|A|) -\frac{a_{22}\gamma-a_{11}}{2\eta} \log\frac{a_{22}\gamma + a_{11}+\eta}{a_{22}\gamma+a_{11}+\eta},
\end{array} \right.
\end{equation}
\emph{where $\mathbf{A}=\sigma^{-2}_n\mathbf{H}'^H\mathbf{H}'+ \mathbf{I}_{2}=\left[  {\begin{array}{*{20}{c}}
  \vspace{-0.2cm} {{a_{11}}}&{{a_{12}}} \\
  {{a_{21}}}&{{a_{22}}}
\end{array}} \right]$ and $\eta=\sqrt {a_{22}^2{\gamma ^2} + 2(2{a_{21}}{a_{12}} - {a_{22}}{a_{11}})\gamma  + a_{11}^2} $. It is easy to find that $\eta$ is a real number since $\mathbf{A}$ is positive definite and $\gamma\geq0$.}

\begin{figure}[t]
  \centering
  \includegraphics[width=8.0cm]{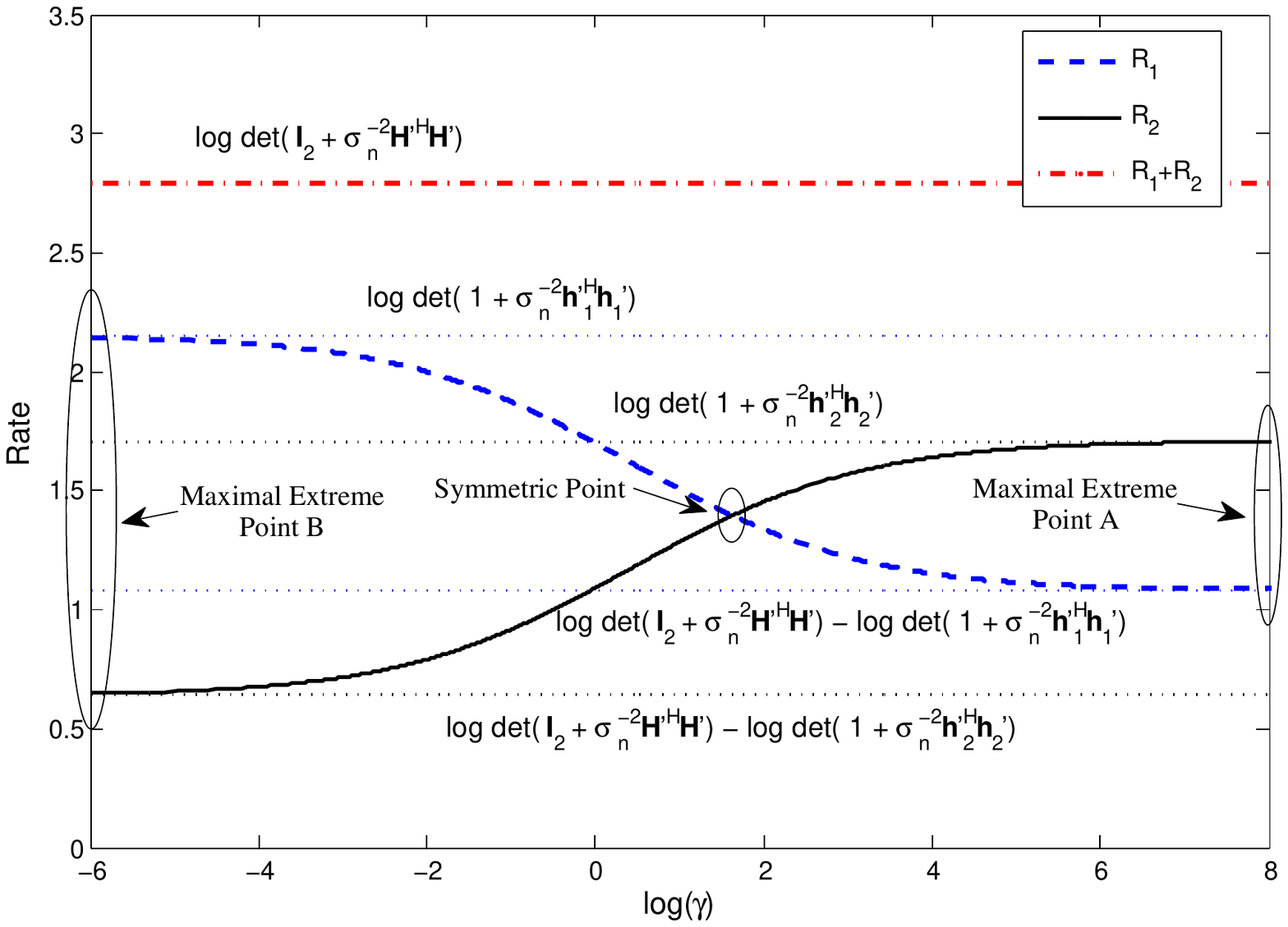}\\\vspace{-0.2cm}
  \caption{Relationship between the user rates and parameter $\gamma$ of the iterative LMMSE detection for two-user MIMO-NOMA system. $N_r=2$, $N_u=2$, $\sigma_N^2=0.5$ and $\mathbf{H}=[
  1.32\;-1.31; \; -1.43 \;0.74]$. }\label{f6}\vspace{-0.4cm}
\end{figure}
It should be noted from (\ref{coro2_1}) that $R_1$ and $R_2$ are non-linear functions of $\gamma$. It is easy to check that $R_1+R_2=\log \det\left( \mathbf{I}_{2} + {\sigma _n^{- 2}}\mathbf{H}'^H \mathbf{H}'\right)$, and when $\gamma\to0$ (or $\gamma\to\infty$), the limit of $(R_1,R_2)$ in (\ref{coro2_1}) converges to the maximal point B (or A) in Fig. \ref{f5}. When the parameter $\gamma$ changes from $0$ to $\infty$, the point $(R_1,R_2)$ can achieve any point on the segment AB in Fig. \ref{f5}. It also shows an alternative proof of \emph{Theorem 3}. In addition, the achievable rates of TDMA and OFDMA are strictly smaller than that of the iterative LMMSE NOMA.

Fig. \ref{f6} presents the rate curves of $R_1$ and $R_2$ respect to the parameter $\gamma$. It verifies that $R_2$ increases monotonously with the $\gamma$ (or $\gamma_2$), and $R_1+R_2$ equals to the sum capacity.\vspace{-0.3cm}

\subsection{MISO: $N_r=1$}
Let $N_r=1$. From (\ref{Eqn:eqv_n}), (\ref{e9}) can be rewritten to
\begin{eqnarray*}
\!\!\!\!\!\!\!\!\!\!\!\!\! &&u_{i,t} \!=\! x_{i,t} \!+\!  \frac{{v}_i^2{{{h}'_i}^H}}{{v}_i\!-\!{v}_{\hat{x}_i}}\!\!\left( \sigma^2_n \!\!+\! \mathbf{h}'{\mathbf{V}_{\bar{\bf{x}}}}\mathbf{h'}^H \right)^{\!-1}\!\!
\left[ \mathbf{h}'\!\left(\mathbf{x}_{\backslash i,t}
\!-\!\bar{\mathbf{x}}_{\backslash i,t}\right) \!+\!\mathbf{n}_t\right]\!\!,\\
\!\!\!\!\!\!\!\!\!\!\!\!\!\!\!\!\!\!\!\!\!\!\!\! &&{v}_{\hat{x}_i}\!=\!{v}_i - {v}_i^2 |{{h}}_i|^2 (\sigma^2_n+{\mathbf{h}'} {\mathbf{V}_{\bar{\bf{x}}}} {\mathbf{h}'}^H)^{-1}.
\end{eqnarray*}
Thus,
\begin{eqnarray}
\!\!\!\!\!\!\!\!\!\!\!\!\!\!\!\!\!\!\!\!\!\!\!\!&&u_{i,t} = x_{i,t} +  \frac{{{h'}_i^H}}{|{{h}'_i}|^2 }\!
\left[ \mathbf{h}'\!\left(\mathbf{x}_{\backslash i,t}
-\bar{\mathbf{x}}_{\backslash i,t}\right) \!+\!\mathbf{n}_t\right],\\
\!\!\!\!\!\!\!\!\!\!\!\!\!\!\!\!\!\!\!\!\!\!\!\!&&\rho_i^{-1}= [{v}_{\hat{x}_i}^{-1}-{v}_i^{-1}]^{-1}= \frac{1}{|h'_i|^2}{\Big[\sum\limits_{k\neq i}|h'_k|^2{v}_k+\sigma^2_n\Big]}.
\end{eqnarray}
Equivalently, it can be rewritten to
\begin{eqnarray}
\!\!\!\!\!\!\!\!\!\!\!\!\!\!\!\!\!\!\!\!\!\!\!\!&& \mathbf{u}_{t} =  \mathbf{\Lambda}_{\mathbf{h'}^H\mathbf{h}'}^{-1}[{\mathbf{h}'}^H{y}_t - \mathbf{\Omega}_{\mathbf{h'}^H\mathbf{h}'}\bar{\mathbf{x}}_t],\label{Eqn:MISOa}\\
\!\!\!\!\!\!\!\!\!\!\!\!\!\!\!\!\!\!\!\!\!\!\!\!&&\mathbf{v}^e=\mathbf{\rho}.^{-1} = (\sigma^2_n+\mathbf{h}'{\mathbf{V}_{\bar{\bf{x}}}}\mathbf{h'}^H) |\mathbf{h}'|.^{-2} -\bar{\mathbf{v}},\label{Eqn:MISOb}
\end{eqnarray}
where $\mathbf{\Lambda}_{\mathbf{A}}={\mathrm{diag}}\{\mathbf{A}\}$,  $\mathbf{\Omega}_{\mathbf{A}}=\mathbf{A}-\mathbf{\Lambda}_{\mathbf{A}}$, and $|\mathbf{h}'|.^{-2}=[|h'_1|^{-2},\dots,|h'_{N_u}|^{-2}]$.

\emph{\underline{Relation to ESE in IDMA:}} Note that \eqref{Eqn:MISOa} and \eqref{Eqn:MISOb} are the same as the ESE in IDMA \cite{Ping2003_1}, which means that the ESE in IDMA is a kind of LMMSE receiver. This explains that IDMA is a good multiple access scheme, since it can achieve the sum capacity of the MISO system.

\subsection{SIMO: $N_u=1$}
Let $N_u=1$. From (\ref{Eqn:eqv_n}), (\ref{e9}) can be rewritten to
\begin{align}
{{\hat {{x}}}_t}
&= {v}_{\hat{x}}\left[{v}^{-1}\bar{{x}}_t+ \sigma_n^{-2}\mathbf{h'}^H\mathbf{y}_t  \right],\\
{v}_{\hat{x}}& = [\sigma_n^{-2}\|\mathbf{h}'\|^2+{v}^{-1}]^{-1},
\end{align}
and
\begin{align}
u_t =  \frac{\mathbf{h'}^H\mathbf{y}_t }{\|\mathbf{h}'\|^2}, \;\;{v}_{\hat{x}} = \frac{\sigma^{2}}{\|\mathbf{h}'\|^2}.\label{SIMO}
\end{align}
In this case, the iteration between the detector and decoders are trivial.

\emph{\underline{Relation to MRC:}} Note that \eqref{SIMO} is the exact MRC \cite{Brennan1959}, which means that MRC is a kind of LMMSE receiver. This shows that MRC is optimal and can achieve the capacity of the SIMO system.

\section{Practical Multiuser Code Design for MIMO-NOMA}

Recently, Low-Density Parity-Chek (LDPC) codes are optimized to support much higher sum spectral efficiency and user loads for MISO in \cite{YHu2018, JSong2018, YZhang2018}. In addition, a LDPC code concatenated with a simple repetition code is constructed to obtain a near MISO capacity performance in \cite{XWang2018}, \cite{XWang20182}.  To further support massive users, an IRA code parallelly concatenated with a repetition code is proposed in \cite{Song-ISIT2015, SongTVT}. However, these design methods do not consider the effect of multiple receive antennas. In this paper, a kind of multi-user IRA code consisting of repetition code and IRA code is optimized for MIMO-NOMA. For more details, please refer to \cite{YC2018}. We will show that the optimized IRA can approaching the MIMO-NOMA capacity (e.g. BER performances are within 0.8dB away from the Shannon limit) for various of system loads. In this section, we give the multi-user IRA code design in detail.

To design suitable multiuser codes for the LMMSE detection, we first derive a transformation between the input-output variance of LMMSE detection and the input-output mutual information of the single-user decoders. Then, based on the EXIT analysis~\cite{Brink2001,Song-ISIT2015, SongTVT,Song-MaxSum}, code parameters can be optimized to match well with LMMSE detection.

To be specific, since the output of LMMSE can be equivalent to the observation from AWGN channel, the \emph{extrinsic} variance associated with the estimated signal from LMMSE is the variance of equivalent noise, such that the \emph{a-priori} mutual information for the decoder is obtained by exploiting the EXIT analysis. For general linear block codes, the EXIT functions can be obtained easily~\cite{Brink2001,Song-ISIT2015, SongTVT,Song-MaxSum}. For the opposite direction, the \emph{a-priori} variance of LMMSE is determined by the \emph{extrinsic} mutual information from the decoder. The whole iterative process will stop when the decoding is successful or the maximum iteration number is reached. In other words, we can statistically trace the iterative message update between LMMSE detection and a bank of single-user decoders. The detailed process is as follows.
\begin{figure*}[t]\vspace{-0.0cm}
      \centering
      \includegraphics[width=15cm]{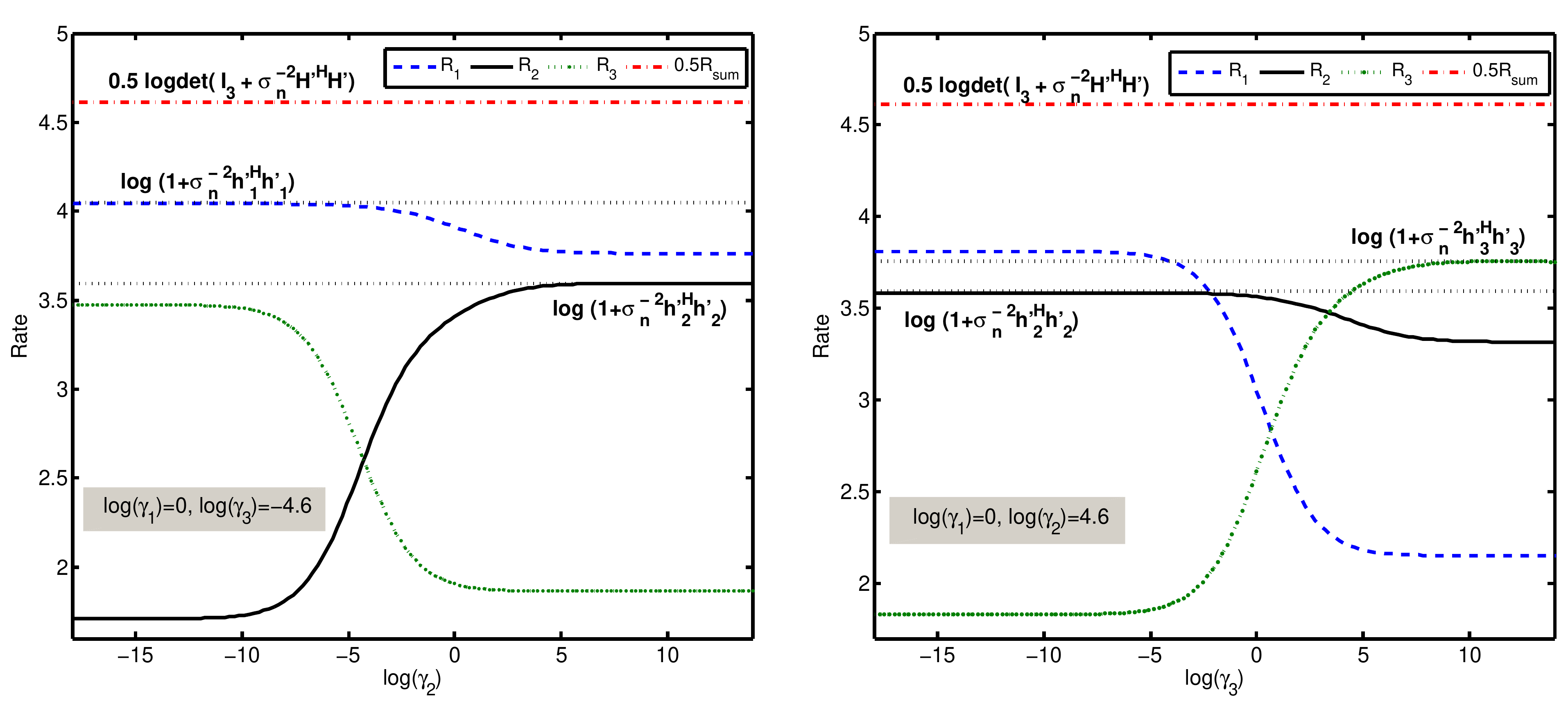}\\\vspace{-0.2cm}
      \caption{Relationship between the user rates and parameters $[\gamma_2,\gamma_3]$ of the iterative LMMSE detection for three-user MIMO-NOMA. $N_r=2$, $N_u=3$, $\sigma_N^2=0.5$ and $\mathbf{H}=[0.678 \;  0.603 \;  0.655;\; 0.557 \;  0.392  \;  0.171]$.}\label{f8}\vspace{-0.0cm}
\end{figure*}
\subsection{LMMSE $\to$ Decoder}
For simplicity, we assume $\mathbf{H}'$ is IID Gaussian, and consider the detection of user $k$. Let $\bar{x}_k$ and $u_k$ be \emph{a-priori} and \emph{extrinsic} estimations of LMMSE detection associated with $x_k$. Correspondingly, let $v_k$ and $v^e_k$ be the variances of $\bar{x}_k$ and $u_k$ respectively. We can obtain the \emph{a-posteriori} output variance ${v}_{\hat{x}_k}$ of LMMSE is \cite{Lei2016,Lei2015,Lei20162}
\begin{equation*}
{v}_{\hat{x}_k}\!=\!
\tiny{\frac{\sqrt{\!(snr^{-1}\!\!+\!N_r\!-\!N_u)^2\!+\!4N_usnr^{-1}}\! -\!(snr^{-1}\!+\!N_r\!-\!N_u)}{2N_u(v_k)^{-1}}\!,}\vspace{-0.1cm}
\end{equation*}\vspace{-0.1cm}
where $snr$$=$${v_k}/{\sigma_n^2}$. \emph{Extrinsic} output variance of LMMSE is
\begin{small}
\begin{align} \nonumber
&v^{e}_k=[(\hat{v}_k)^{-1}-(v_k)^{-1}]^{-1} \\ \nonumber
&=(v_k)\frac{\sqrt{(snr^{-1}\!+\!N_r\!-\!N_u)^2 \!+\!4N_usnr^{-1}}\!-(snr^{-1}\!+\!N_r\!-\!N_u)}{(snr^{-1}\!\!+\!N_r\!+\!N_u) \!-\!\sqrt{(snr^{-1}\!+\!N_r\!-\!N_u)^2\!+\!4N_usnr^{-1}}}
\end{align}
\end{small}
\!\!Based on Proposition 1, we can rewritten $u_k=x_k+\tilde{z}_k$, where $\tilde{z}_k$ is an equivalent Gaussian noise with mean $0$ and variance $Var(\tilde{z}_k)=Var(u_k)=v^e_k$. Therefore, \emph{a-priori} mutual information associated with $x_k$ for the DEC can be obtained.

\subsection{Code Optimization $\to$ Detector}
Following the similar methods in~\cite{Song-MaxSum,Song-ISIT2015, SongTVT}, the EXIT function of repetition-aided IRA can be obtained and then \emph{extrinsic} mutual information $I_k^e$ is calculated. According to EXIT analysis~\cite{Brink2001,Song-MaxSum,Song-ISIT2015, SongTVT}, output log-likelihood ratio $L^e_k$ obeys Gaussian distribution $\mathcal{N}((J^{-1}(I_k^e))^2/2, (J^{-1}(I_k^e))^2)$, where function $J(\cdot)$ is given in~\cite{Brink2001}. Since $x_k$ is a BPSK signal, variance $v_k=E_{L_k^e}[1-({\rm{tanh}}(L_k^e/2))^2]$ is obtained by Monte Carlo simulations, which is fed back to the LMMSE.

By using this variance-EXIT transfer process between the LMMSE and decoder, we trace statistically the message update and then optimize the parameters of repetition-aided IRA codes to match well with the LMMSE.\vspace{-0.1cm}

\section{Numerical Results}
This section presents the numerical results of achievable rate of three-user MIMO-NOMA, and provides the BER simulations for the proposed iterative LMMSE detection with optimized multi-user codes.

\begin{figure}[t]
      \centering
      \includegraphics[width=9.0cm]{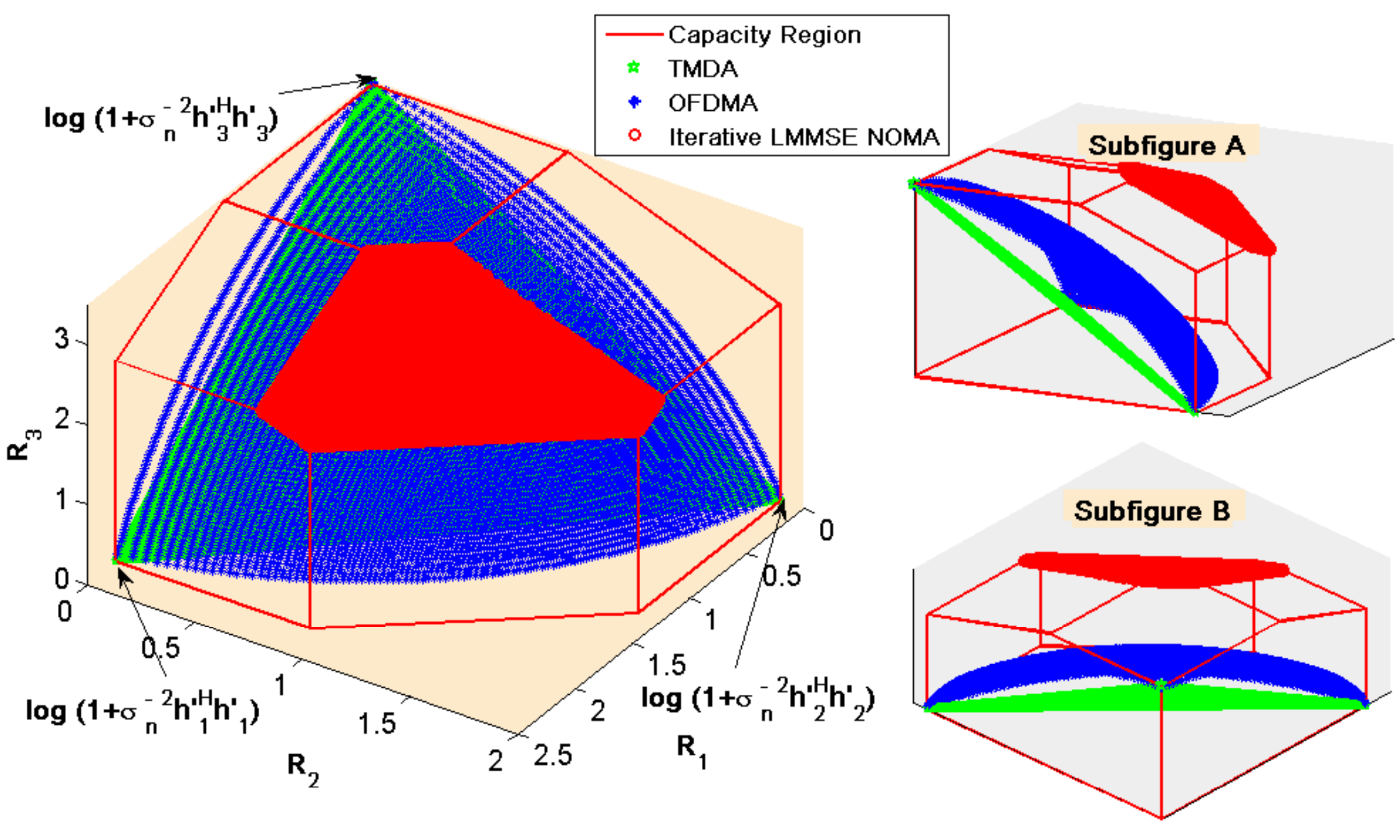}\\
      \caption{Achievable rates for all $(\gamma_1,\gamma_2)$ of the iterative LMMSE detection for three-user MIMO-NOMA system. $N_r=3$, $N_u=3$, $\sigma_N^2=0.5$ and $\mathbf{H}=[1.95\;  1.28 \;  -2.53;\;
                 -0.31 \;  -0.16  \;  2.22;\; 0.55\; 1.08\; -1.98]$. Subfigure A and Subfigure B are the same figure with different rotated viewports.}\label{f9}\vspace{-0.4cm}
\end{figure}

\begin{figure*}[t]\vspace{-0.0cm}
      \centering
      \includegraphics[width=15cm]{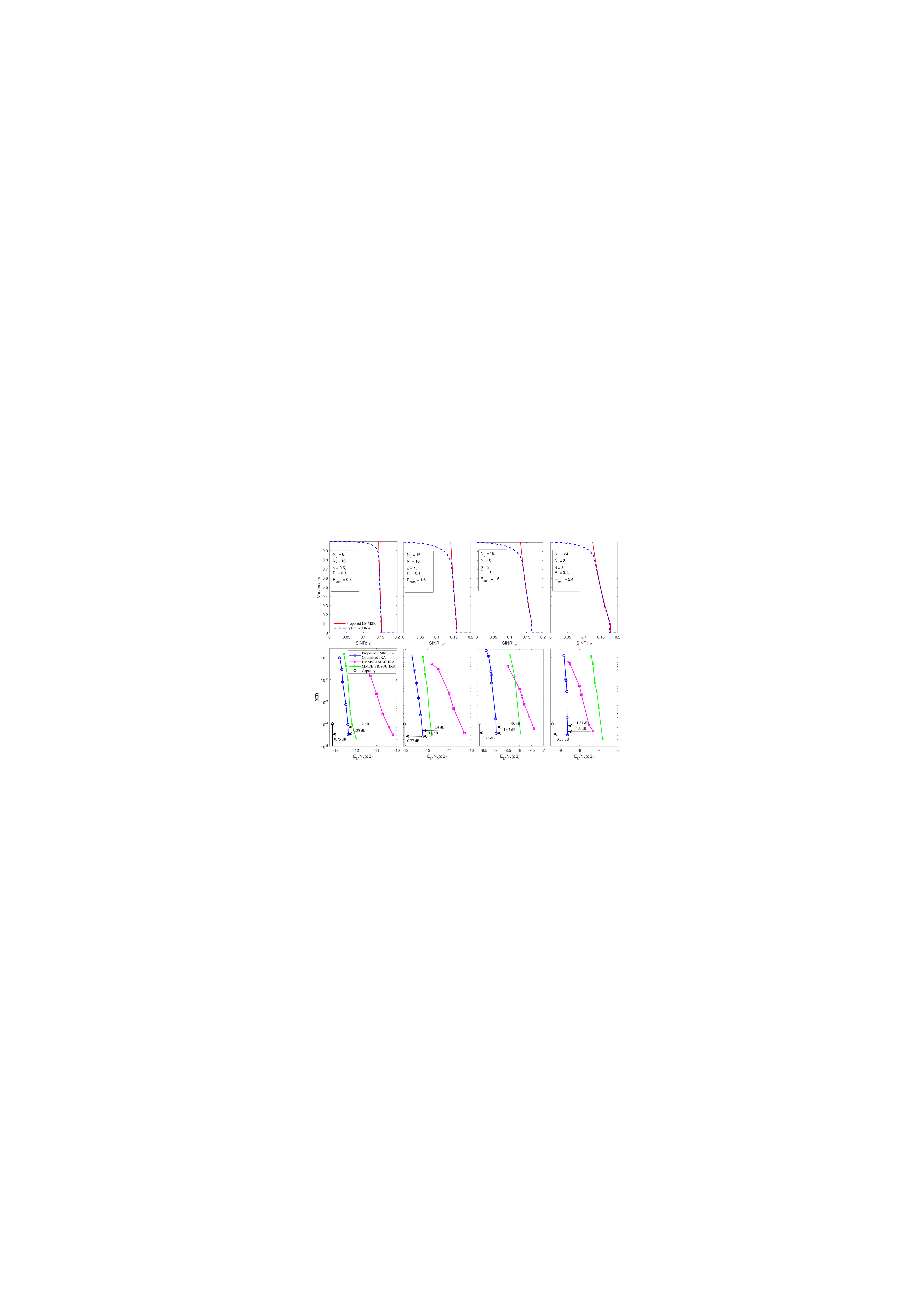}\\\vspace{-0.2cm}
      \caption{SINR-variance transfer charts and BER performances of the LMMSE Receiver for MIMO-NOMA with channel load $\beta=\{0.5, 1, 2, 3\}$, where user number $N_u$ and receive antenna $N_r$ are $(N_u, N_r)=(8, 16), (16, 16), (16, 8), (24, 8)$ respectively. Each user is encoded by an optimized IRA code with code rate 0.1 bits/symbol and code length $4.096\times 10^4$. The use rate of MAC-IRA code is $0.08$ and decoding threshold is $0.03$~dB from the MAC capacity. The rate of SU-IRA is $0.1$~and decoding threshold is from $0.05$~dB from the single-user capacity.}\label{EXIT_BER}\vspace{-0.1cm}
\end{figure*}
\subsection{Three-user MIMO-NOMA}
For three-user MIMO-NOMA, it is hard to get a closed-form solution of the user rates. Hence, it is difficult to show the exact achievable rate region of the iterative LMMSE detection. However, the user rates in (\ref{lemma1}) can be solved numerically.

Fig. \ref{f8} shows the relationships between the user rates and $[\gamma_2,\gamma_3]$ with $\gamma_1=1$, where $N_r=2$, $N_u=3$, $\sigma_N^2=0.5$, and $\mathbf{H}=[0.678 \;  0.603 \;  0.655;\; 0.557 \;  0.392  \;  0.171]$. Notice that although the user rates change with $\gamma_2$ and $\gamma_3$, the sum rate $R_{sum}$ is constant and equals to the system sum capacity. Furthermore, the user rate $R_2$ increases monotonously with $\gamma_2$, but $R_1$ and $R_3$ decrease monotonously with $\gamma_2$. Similarly, the user rate $R_3$ increases monotonously with $\gamma_3$, but $R_1$ and $R_2$ decrease monotonously with $\gamma_3$.

In Fig. \ref{f9}, the system capacity region is the polygonal consisted by the red lines, which is dominated by the red hexagonal face. The red points in Fig. \ref{f9} are the achievable points of the iterative LMMSE detection. It shows that as we change the values of $\gamma_2$ and $\gamma_3$, the achievable points can reach any point on the dominated hexagonal face. Therefore, for the three-user MIMO-NOMA, the iterative LMMSE detection can also achieve all points in the capacity region, i.e., the iterative LMMSE detection is an optimal detection. In addition, we can see that the achievable rates of TDMA and OFDMA are strictly smaller than that of the iterative LMMSE NOMA.

It should be noted that the results in this paper can also apply to the overloaded MIMO-NOMA systems (like Fig. \ref{f8}) that the number of users is larger than the number of BS antennas, i.e., $N_u>N_r$.\vspace{-0.4cm}

\subsection{BER Performance with Optimized IRA Codes}

Here, we assume that each user employs a repetition-aided IRA code proposed for the Multiple-Access Channel (MAC)~\cite{Song-ISIT2015, SongTVT}, which is constructed by parallelly concatenating a repetition code and IRA code. In this paper, we optimize the repetition-aided IRA codes over MIMO-NOMA systems with channel load $\beta=\{0.5, 1, 2, 3\}$, where user number $N_u$ and receive antenna $N_r$ are $(N_u, N_r)=(8, 16), (16, 16), (16, 8)$, and $(24, 8)$, respectively. The corresponding optimized code parameters are given in Table~\ref{Opt_degree1}, which illustrates that these decoding thresholds are very close to the Shannon limits.

To verify the finite-length performance of the repetition-aided IRA codes, we provide the BER performances of the optimized codes. Each user employs a random interleaver and the length of information vector for each user is $4096$. The rate of each user is $R_u=0.1$ bits/symbol, and the sum rate is $R_{sum}=0.1*N_u$ bits per channel use. $E_b/N_0$ is calculated by $E_b/N_0=\frac{P_u}{2R_u\sigma^2_n}$, where $P_u=1$ is the power of each user, and $\sigma^2_n$ is the variance of the Gaussian noise. The standard sum-product algorithm is used for the single-user decoding, in which the maximum iteration number is $250$. Fig.~\ref{EXIT_BER} shows that for all $\beta$, gaps between the BER curves of the codes at $10^{-5}$ and the corresponding Shannon limits are about $0.7 \sim 0.8$~dB.
\begin{table}[t]\vspace{-0.0cm}
\caption{Optimized Repetition-aided IRA codes over MIMO-NOMA}\label{Opt_degree1}
\centering
\begin{tabular}{|c|c|c|c|c|}
\hline
$\it{\beta}$ & {$0.5$} & {$1$}& {$2$} & {$3$} \\
\hline
$\textit{$N_u$}$ & {$8$} & {$16$}& {$16$} & {$24$} \\
\hline
$\textit{$N_r$}$ & {$16$} & {$16$}& {$8$} & {$8$} \\
\hline
${\textit{$R_u$}}$ & $0.1$  & $0.1$  & $0.1$ & $0.1$\\
\hline
${\textit{N}}$ & $4\times10^{4}$  & $4\times10^{4}$  & $4\times10^{4}$ & $4\times10^{4}$\\
\hline
${\textit{R}}_{\text{sum}}$ & $0.8$  & $1.6$  & $1.6$ & $2.4$\\
\hline
${\textit{q}}$ & $1$  & $2$  & $2$ & $2$\\
\hline
${\it{\alpha}}$ & $2$  & $2$  & $2$ & $2$ \\
\hline
${\it{\lambda}}_{\text{3}}$ & $0.087105$ & $0.1016$ & $0.107994$ & $0.116863$\\
${\it{\lambda}}_{\text{10}}$ & $0.138217$ & $0.138386$ & $0.129009$ & $0.127289$\\
${\it{\lambda}}_{\text{30}}$ &  $0.207022$ & $0.262982$ & $0.219708$ &$0.159387$\\
${\it{\lambda}}_{\text{80}}$ &  $0.068682$   & $0.114347$ &  $0.141601$ & $0.234121$\\
${\it{\lambda}}_{\text{100}}$ &  $0.498975$ & $0.382685$ & $0.401687$ & $0.36234$ \\
\hline
$(\frac{\textit{E}_{\textit{b}}}{\textit{N}_{\text{0}}})^{\it{\ast}}_{\text{dB}}$ & $ -13.14$ & $-12.95$ & $-9.66$ & $-9.35$ \\
\hline
$\text{S.~L.}$ &  $ -13.16$ &  $-13.03$  &  $-9.7$  &  $-9.38$ \\
\hline
\end{tabular}
\end{table}

To validate the advantage of the proposed system through matching between LMMSE detector and optimized IRA codes, we provide two state-of-art systems for comparisons, which are LMMSE detector combined with an existing repetition-aided IRA code~\cite{Song-ISIT2015, SongTVT}, and MMSE-SIC detector~\cite{Wang1999,Studer2011} combined with a capacity-approaching Single-User IRA (SU-IRA) code. Note that the parameters of repetition-aided IRA code~\cite{Song-ISIT2015, SongTVT} are $\lambda(x)=0.063021x+0.228288x^2+0.111951x^{9}+0.226877x^{29}+0.369864x^{49}$, $q=5$, and $\alpha=1$, denoted as MAC-IRA code, whose rate is $0.08$ and decoding threshold is $0.03$~dB from the MAC capacity. The parameters of SU-IRA are $0.085867x^2+0.132226x^9+0.198883x^{29}+0.276011x^{79}+0.307013x^{99}$, $q=1$, and $\alpha=2$, whose rate is $0.1$~and decoding threshold is from $0.05$~dB from the single-user capacity. As shown as Fig.~\ref{EXIT_BER}, when the BER curves of three systems are at $10^{-4}$, the optimized IRAs have $1.4 \sim 2$~dB performance gains over the un-optimized IRAs, and $0.38 \sim 1.3$~dB performance gains over the systems consisting of MMSE-SIC detector and the SU-IRA code. These comparisons demonstrate that multiuser code optimization provides a promising new treatment for the applications of MIMO-NOMA technologies.

\section{Conclusion}
The theoretical limit of the PIC iterative receiver has been an open problem for a long time, especially for the multi-user MIMO channel. This paper analyzes the achievable rate region of the iterative LMMSE multi-user detection for both symmetric and asymmetric MIMO-NOMA. For the symmetric case, it is proved that iterative LMMSE detection achieves the capacity of MIMO-NOMA with any number of users; while for the asymmetric case, it is proved that the iterative LMMSE detection achieves the \emph{sum capacity} of MIMO-NOMA with any number of users. In addition, all the maximal extreme points in the capacity region of MIMO-NOMA with any number of users are achievable, and all points in the capacity regions of two-user and three-user systems are also achievable. Finally, a kind of IRA multiuser code is designed for the iterative LMMSE receiver. Simulation results show that under different channel loads, the BERs of the proposed iterative LMMSE detection are within 0.8dB from the Shannon limits and outperform the existing methods. Furthermore, the improvement is more notable for large system overloads (e.g. $\beta\geq3$), while for small system overloads (e.g. $\beta\leq0.5$), the AWGN SU-IRA and the MMSE SIC with SU-IRA is good enough since the user interference is negligible.

How to design a low-complexity iterative receiver to achieve the capacity region of the general vector multiple access channel \cite{YuW2004} will be an interesting future work.
\appendices
\section{Derivation of \emph{A-Posteriori} LMMSE}\label{APP:LMMSE}
We assume $\mathbf{x}(t)\sim \mathcal{CN}(\bar{\mathbf{x}}(t),\mathbf{V}_{\bar{\mathbf{x}}})$, i.e. $p(\mathbf{x}(t)) \varpropto e^{-{(\mathbf{x}(t)-\bar{\mathbf{x}}(t))^H \mathbf{V}^{-1}_{\bar{\mathbf{x}}}(\mathbf{x}(t)-\bar{\mathbf{x}}(t))}}$. Since $\mathbf{n}(t)\sim \mathcal{CN}(\mathbf{0},\sigma_n^2\mathbf{I})$, we have $p(\mathbf{y}_t|\mathbf{x}(t))\varpropto e^{-\frac{(\mathbf{y}_t-\mathbf{H'x}(t))^H(\mathbf{y}_t-\mathbf{H'x}(t))}{\sigma_n^2}}$.
Thus, the \emph{a-posteriori} conditional probability of $\mathbf{x}(t)$ given $\mathbf{y}_t$ is
\begin{eqnarray}
\!\!\!\!\!\!\!\!\!\!\!\!\!\!\!\!\!&p&\!\!\!\!\!\! (\mathbf{x}(t)|\mathbf{y}_t)\nonumber\\
\!\!\!\!\!\!\!\!\!\!\!\!\!\!\!\!\!&=& \!\!\!\!p(\mathbf{x}(t))p(\mathbf{y}_t|\mathbf{x}(t)) \nonumber
\\ \!\!\!\!\!\!\!\!\!\!\!\!\!\!\!\!\!&\varpropto&\!\!\!\!\!  e^{-\mathbf{x}(t)^H \left[\sigma _{{{n}}}^{- 2}\mathbf{H}'^H\mathbf{H}'+\mathbf{V} _{{{ \bar{\mathbf{x}}}}}^{-1}\right]\mathbf{x}(t) +2\mathbf{x}(t)^H\left[\mathbf{V}_{\bar{\mathbf{x}}}^{-1}\bar{\mathbf{x}}(t)+ \sigma^{-2}_n\mathbf{H}'^H\mathbf{y}_t  \right]}\nonumber \\
\!\!\!\!\!\!\!\!\!\!\!\!\!\!\!\!\!&\varpropto& \!\!\!\!e^{-\mathbf{x}(t)^H { \mathbf{V} _{{{\hat {\mathbf{x}}}}}^{-1}}\mathbf{x}(t) +2\mathbf{x}(t)^H{{ \mathbf{V} _{{{\hat {\mathbf{x}}}}}^{-1}} \hat {\mathbf{x}}(t)}}
\end{eqnarray}
Therefore, the \emph{a-posteriori} estimation and variance are
\begin{eqnarray}
 &{{\hat {\mathbf{x}}}(t)}&
= \mathbf{V}_{\hat {\mathbf{x}}}\left[\mathbf{V}_{\bar{\mathbf{x}}}^{-1}\bar{\mathbf{x}}(t)+ \sigma^{-2}_n\mathbf{H}'^H\mathbf{y}_t  \right],\\
&\mathbf{V} _{{{\hat {\mathbf{x}}}}}& = (\sigma _{{{n}}}^{- 2}\mathbf{H}'^H\mathbf{H}'+\mathbf{V} _{{{\bar{\mathbf{x}}}}}^{-1})^{-1}.
\end{eqnarray}
Hence, we obtain \eqref{GMP2}.

\section{Proof of Proposition 1}\label{ASS3_AP}
The \emph{a-posteriori} LMMSE in Eq. (\ref{GMP2}) can be rewritten to
\begin{equation*}
{{\hat {\mathbf{x}}}(t)}
= \bar{\mathbf{x}}(t) + V_{\bar{\mathbf{x}}}\mathbf{H}'^H\left(\sigma_n^2\mathbf{I}_{N_r}+ \mathbf{H}'V_{\bar{\mathbf{x}}}\mathbf{H}'^H \right)^{-1}\left(\mathbf{y}_t-\mathbf{H}'\bar{\mathbf{x}}(t)\right).
\end{equation*}
From (\ref{e9}), we get $u_{i,t} = x_{i,t} + n^{*}_{i,t},$
and
\begin{eqnarray} \nonumber
n_{i,t}^{*}=  \frac{v_i}{v_{\hat{x}_i}\phi_i} {\mathbf{h}'_i}^H\!\!\left( \sigma^2_n\mathbf{I}_{N_r}\!\!+\! \mathbf{H}'\mathbf{V}_{\bar{\mathbf{x}}}\mathbf{H}'^H \right)^{\!-1}\cdot\quad\quad\\
\left[ \mathbf{H}'\!\left(\mathbf{x}_{\backslash i}(t)
-\bar{\mathbf{x}}_{\backslash i}(t)\right) \!+\!\mathbf{n}(t)\right],\label{Eqn:eqv_n}
\end{eqnarray}
where $\mathbf{x}_{\backslash i}(t)$ (or $\bar{\mathbf{x}}_{\backslash i}(t)$) denotes the vector whose $i$th entry of $\mathbf{x}(t)$ (or $\bar{\mathbf{x}}(t)$)  is set to zero. The equivalent noise $n_{i,t}^{*}$ is independent of $x_{i,t}$. In Eq. (21) of \cite{Rangan2016} and Theorem 4(b) of \cite{Takeuchi2017}, a rigorous proof is elaborated to show that $n_{i,t}^{*}$ is Gaussian distributed, i.e., $n_{i,t}\sim \mathcal{CN}\left(0,1/\phi_i(\mathbf{v}_{\bar{\mathbf{x}}})\right)$. Hence, we obtain
the proposition.

\section{Proof of Lemma 1}\label{APP_Lemma1}

From (\ref{e35}), the achievable rate of user $i$ is given by
\begin{eqnarray}\label{e47}
\!\!\!\!\!\!R_i\!\!\!\!\!\!&=&\!\!\!\!\!\!\int\limits_0^\infty  {{{\left( {{\rho _i} + {\psi _i}{{({\rho _i})}^{- 1}}} \right)}^{- 1}}d{\rho _i}}\nonumber\\
&\mathop \leq \limits^{(a)}&\!\!\!\!\!\!\int\limits_{\phi_i(1)}^{\phi_i(0)}  {{\left[ {\rho_i } +  \left( {\phi_i }^{- 1}{({\rho_i })}\right)^{- 1} \right]}^{ - 1}d{\rho_i }} \!+\!\!\! \int\limits_{0}^{\phi_i(1)} {(1+\rho_i)^{-1}d\rho_i} \nonumber\\
&\mathop = \limits^{(b)}&\!\!\!\!\!\!\int\limits_{v_i=1}^{v_i=0}  {\left( v_i^{-1} +  \phi_i(v_i) \right)^{- 1}d{\phi_i(v_i) }} + \log\left(1+\phi_i(v_i)\right) \nonumber\\
&\mathop = \limits^{(c)}&\!\!\!\!\!\!\int\limits_{v_i=1}^{v_i=0} \!\!\!\! { v_{\hat{x}_i}(v_i) d {v_{\hat{x}_i}(v_i)}^{-1}} \!-\!\!\!\! \int\limits_{v_i=1}^{v_i=0} \!\!\!\! {v_{\hat{x}_i}(v_i)d{v_i^{-1} }} \!\!- \log v_{\hat{x}_i}(\!v_i\!=\!1\!) \nonumber\\
&\mathop = \limits^{(d)}&\!\!\!\!\!\! - \int\limits_{v_1=1}^{v_1=0} { \gamma_i^{-1} \left[\mathbf{V}_{\hat{\mathbf{x}}}(v_1)\right]_{i,i} dv_1^{-1}}
-\mathop {\lim }\limits_{v_1 \to 0} \; \log\left[\mathbf{V}_{\hat{\mathbf{x}}}(v_1)\right]_{i,i}\;\nonumber\\
&\mathop = \limits^{(e)}&\!\!\!\!\!\! \int\limits_{v_1=1}^{v_1=0} \left[v_1- \gamma_i^{-1} \left[\mathbf{V}_{\hat{\mathbf{x}}}(v_1)\right]_{i,i} \right] dv_1^{-1} - \log(\gamma_i).
\end{eqnarray}
The inequality $(a)$ is derived by (\ref{em1})$\sim$(\ref{em3}) and the equality holds if and only if there exists such a code whose transfer function satisfies the matching conditions. The equations $(b)\sim(d)$ are given by $\rho_i=\phi_i(v_i)$, (\ref{ephi}) and (\ref{ev_1}), equation $(e)$ comes from  (\ref{e46}) and (\ref{evall}). In APPENDIX \ref{APP:Code_existence}, we show the existence of such codes whose SINR-variance transfer functions match that of the LMMSE detector. In APPENDIX \ref{APP_existence}, the existence of the infinite integral of (\ref{lemma1}) is proven.%

\section{The Code Existence in Lemma 1}\label{APP:Code_existence}
We first introduce an important property that is established in \cite{Yuan2014}, which builds the relationship between the code rate and its transfer function $\psi_i(\rho_i)$.

\emph{Property of SCM Code}: \emph{Assume $\psi(\rho)$ satisfies }\vspace{0.2cm}\\
\noindent
\hangafter=1
\setlength{\hangindent}{2em}\emph{(i) $\psi(0)=1$ and $\psi(\rho)\geq 0$, for $\rho\in[0,\infty);$}\\
\noindent
\hangafter=1
\setlength{\hangindent}{2em}\emph{(ii) monotonically decreasing in $\rho\in[0,\infty)$;}\\
\noindent
\hangafter=1
\setlength{\hangindent}{2em}\emph{(iii) continuous and differentiable in $[0,\infty)$ except for a countable set of values of $\rho$;}\\
\noindent
\hangafter=1
\setlength{\hangindent}{2em}\emph{(iv) $\mathop {\lim }\limits_{{\rho } \to \infty } {\rho }{\psi }({\rho }) = 0$.}\vspace{0.2cm}

\emph{$\!\!\!\!\!\!$Let $\Gamma_n$ be an $n$-layer SCM code with \emph{SINR-variance} transfer function $\psi^n(\rho)$ and rate $R_n$. Then, there exists $\{\Gamma_n\!\}$ such that: (i) $\psi^n(\rho)\!\leq\!\psi(\rho), \forall\rho\!\geq\!0, \forall n$; (ii), ${R_n} \to R \left(\psi(\rho)\right)$ as $n\!\to\! \infty$, where $R \left(\psi(\rho)\right)$ denotes code rate of transfer function $\psi(\rho)$.}

This property means that there exists such an $n$-layer SCM code $\Gamma_n$ whose transfer function can approach $\psi(\rho)$ that satisfies the conditions (i)$\sim$(iv) with arbitrary small error when $n$ is large enough.

From the ``\emph{Property of SCM Codes}", we can see that there exist such $n$-layer SCM codes whose transfer function satisfies (i)$\sim$(iv) when $n$ is large enough. Therefore, it only needs to check the matched transfer function meets the conditions (i)$\sim$(iv) in order to show the existence of such codes. It is easy to see that conditions (i) and (iv) are always satisfied by (\ref{em1}) and (\ref{em2}) respectively. From (\ref{ephi})$\sim$(\ref{em3}), we can see that $\psi_i(\rho_i)$ is continuous and differentiable in $[0,\infty)$ except at $\rho_i=\phi_i(0)$ and $\rho_i=\phi_i(1)$. Thus, Condition (iii) is satisfied. To show the monotonicity of the transfer function, we first rewrite (\ref{e38}) by the random matrix theorem as
\begin{eqnarray}\label{er2}
\!\!\!\!\phi_i(v_i)\!\!\!\!&=&\!\!\!\!\!\! {{{{\left[ {v_i \!-\! {v_i^2}\frac{{{w^2}}}{{\sigma _n^2}}\mathbf{h}_i^H{{\left( {{{\mathbf{I}}_{{N_r}}} \!\!+\! \frac{{{w^2}v_i}}{{\sigma _n^2}}{\mathbf{H}}{{\mathbf{H}}^H}} \right)}^{\!-\! 1}}{\!\!\mathbf{h}_i}} \right]}^{\!-\! 1}} \!\!- {v_i^{- 1}}} } \nonumber\\
\!\!\!\!&=&\!\!\!\!\!\!
  {\left[ {{{\left( {\frac{{{w^2}}}{{\sigma _n^2}}\mathbf{h}_i^H{{\left( {{v_i^{ - 1}}{{\mathbf{I}}_{{N_r}}} \!+\! \frac{{{w^2}}}{{\sigma _n^2}}{\mathbf{H}}{{\mathbf{H}}^H}} \right)}^{- 1}}{\mathbf{h}_i}} \right)}^{ \!-\! 1}} \!\!\!-\! 1} \right]}^{- 1} \nonumber \\
\!\!\!\!&=&\!\!\!\! {{1 \mathord{\left/
 {\vphantom {1 {\left( {f_i^{- 1}(v_i) - 1} \right)}}} \right.
 \kern-\nulldelimiterspace} {\left( {f_i^{- 1}(v_i) - 1} \right)}}},
\end{eqnarray}
where $f_i(v_i)={\frac{{{w^2}}}{{\sigma _n^2}}\mathbf{h}_i^H{{\left( {{v_i^{- 1}}{{\mathbf{I}}_{{N_r}}} + \frac{{{w^2}}}{{\sigma _n^2}}{\mathbf{H}}{{\mathbf{H}}^H}} \right)}^{- 1}}{\mathbf{h}_i}}$. It is easy to check that $f_i(v_i)$ is a decreasing function with respect to $v_i$, and $\phi_i(v_i)$ is thus a decreasing function of $v$. With the definition of $\psi(\rho)$ from (\ref{em1})$\sim$(\ref{em3}), we then see that $\psi_i(\rho_i)$ is a decreasing function in $[0,\infty)$. Therefore, the matched transfer function can be obtained by the SCM code, i.e., there exists such codes that satisfy the matching conditions.
\section{The Existence of Infinite Integral (\ref{lemma1})}\label{APP_existence}
With (\ref{lemma1}), we have
\begin{eqnarray}
{R_i} \!\!\!\!&=&  \!\!\!\!- \int\limits_{{v_1} = 1}^{{v_1} = 0} {\gamma _i^{- 1}{{\left[ {{{\mathbf{V}}_{{\mathbf{\hat x}}}}({v_1})} \right]}_{i,i}}dv_1^{- 1}}  - \mathop {\lim }\limits_{{v_1} \to 0} \;\log ({\gamma _i}{v_1})\nonumber\\
  &\mathop =\limits^{(a)}& \!\!\!\! - \int\limits_0^\infty  {{{\left[ {{{\left( {{\mathbf{A}_{\bm{\gamma}} } + s{{\mathbf{I}}_{{N_u}}}} \right)}^{- 1}}} \right]}_{i,i}}ds}  - \mathop {\lim }\limits_{s \to \infty } \;\log ({\gamma _i}s^{-1})
  \nonumber\\
  &\mathop =\limits^{(b)}& \!\!\!\! - \int\limits_0^\infty  {{\mathbf{u}_i}^H{{\left( {{\bm{\Lambda} _{{A_{\bm{\gamma} } }}} + s{{\mathbf{I}}_{{N_u}}}} \right)}^{- 1}}{\mathbf{u}_i}ds}  - \mathop {\lim }\limits_{s \to \infty } \;\log ({\gamma _i}s^{-1}), \nonumber\\
 &\mathop =\limits^{(c)}& \!\!\!\!  - \int\limits_0^\infty  {\sum\limits_{j = 1}^{{N_u}} {{{\left\| {{u_{ij}}} \right\|}^2}} {{\left( {{\lambda _{{{\mathbf{A}_{\bm{\gamma}, }}}j}} + s} \right)}^{- 1}}ds}  - \mathop {\lim }\limits_{s \to \infty } \;\log ({\gamma _i}{s^{- 1}}) \nonumber\\
  &=& \!\!\!\! \sum\limits_{j = 1}^{{N_u}} {{{\left\| {{u_{ij}}} \right\|}^2}} \log \left( {{\lambda _{{{\mathbf{A}_{\bm{\gamma} }},}j}}} \right) - \log ({\gamma _i}),
\end{eqnarray}
where equation (a) comes from $ s=v_1^{-1}$ and $\mathbf{A}_{\bm{\gamma} } =\bm{\Lambda} _{\bm{\gamma} }^{1/2} \left( {\sigma _n^{- 2}{{{\mathbf{H'}}}^H}{\mathbf{H'}} + {{\mathbf{I}}_{{N_u}}}} \right)\bm{\Lambda} _{\bm{\gamma} }^{1/2}$; equation (b) is based on $\quad \mathbf{A}_{\bm{\gamma} } = \mathbf{U}^H\bm{\Lambda}_{\mathbf{A}_{\bm{\gamma} }} \mathbf{U}$ and $\mathbf{u}_i$ is the $i$th column of $\mathbf{U} $;  ${\lambda _{{{\mathbf{A}_{\bm{\gamma} }},}j}}$ is the $i$th diagonal element of ${\bm{\Lambda} _{{A_{\bm{\gamma} } }}}$. Thus, we show the existence of the infinite integral (\ref{e47}), i.e., $R_i$ has finite value.\vspace{-0.0cm}

\section{Proof of Theorem 1}\label{APP_Theorem2}
With (\ref{lemma1}), the achievable sum rate is
\begin{align}\label{e49}
&R_{sum}=\sum\limits_{i = 1}^{{N_u}} {{R_i}} \nonumber\\
\!\!\!\!\!\!\!&\mathop = \limits^{(a)}\!-\!\!\!\! \int\limits_{v_1=1}^{v_1=0} \!\sum\limits_{i = 1}^{{N_u}} {{\!\big(\!\gamma_i^{-1} \!\left[\mathbf{V}_{\hat{\mathbf{x}}}(v_1)\right]_{i,i} \!\big) dv_1^{-1}}  }
\!-\!\!\mathop {\lim }\limits_{v_1 \to 0}  \log(v_1^{N_u}\mathop \Pi \limits_{i = 1}^{{N_u}} \gamma_i) \nonumber\\
\!\!\!\!\!\!\!&=\!-\!\!\!\!\int\limits_{v_1=1}^{v_1=0}  {{ \mathrm{Tr}\{\bm{\Lambda}_{\bm{\gamma}}^{-1} \mathbf{V}_{\hat{\mathbf{x}}}(v_1) \} dv_1^{-1}}  }
\!-\!\mathop {\lim }\limits_{v_1 \to 0}  \log(v_1^{N_u}\mathop \Pi \limits_{i = 1}^{{N_u}} \gamma_i) \nonumber\\
\!\!\!\!\!\!\!&\mathop = \limits^{(b)}\!-\!\!\mathop {\lim }\limits_{v_1 \!\to 0} \!\log(v_1^{\!N_u}\!\!\mathop \Pi \limits_{i = 1}^{{N_u}} \!\!\gamma_i)\!-\! \!\left[ \log\!| (v_1^{\!-\!1}\!\!\!-\!1\!)\mathbf{I}_{\!N_{\!u}} \!\!\!+\!\! \left( \mathbf{I}_{\!N_{\!u}} \!\!\!+ \!\sigma_n^{\!-\!2}\mathbf{H}'^H\!\mathbf{H}' \right)\!\bm{\Lambda}_{\bm{\gamma}}\! |\right]_{\!v_1\!=1}^{\!v_1\!=0}
\nonumber\\
\!\!\!\!\!\!\!& =\!\!-\!\!\mathop {\lim }\limits_{v_1\! \to 0} \! \log(v_1^{\!N_{\!u}}\!\mathop \Pi \limits_{i \!=\! 1}^{{\!N_{\!u}}}\!\! \gamma_i)\!-\!\!\! \mathop {\lim }\limits_{v_1\! \to 0} \!{\log \!|v_1^{\!-\!1}\!\mathbf{I}_{\!N_{\!u}}\!| } \!+\! \log\! | (\mathbf{I}_{\!N_{\!u}} \!\!\!+\! {\sigma _n^{\!-\! 2}}\mathbf{H}'^H\!\mathbf{H}')\bm{\Lambda}_{\!\bm{\gamma}}|
 \nonumber\\
\!\!\!\!\!\!\!&=  \log | \mathbf{I}_{N_u} + {\sigma _n^{- 2}}\mathbf{H}'^H\mathbf{H}' |,\nonumber
\end{align}
which is the exact system sum capacity of MIMO-NOMA system. Equation $(a)$ is derived by (\ref{lemma1}), and equation $(b)$ is based on (\ref{evall}) and the law $\int {\mathrm{Tr}\{\left(s\mathbf{I} + \mathbf{A}\right)^{-1}\} ds }= \log|s\mathbf{I} + \mathbf{A}|$. It means iterative LMMSE detection is sum capacity-achieving.\vspace{-0.2cm}

\section{Capacity Region Domination Lemma}\label{CR_domination_lemma}
The following lemma is used of the proofs in the rate analyses of iterative LMMSE detection.

{\emph{Capacity Region Domination Lemma \cite{Han1979}}}: \emph{All the points in the capacity region $\mathbf{\mathcal{R}}_\mathcal{S}$ is dominated by a convex combination of the following $(N_u!)$ \textbf{maximal extreme points}.}
\begin{equation}\label{dominate}
\left\{\begin{array}{l}
{R_{{k_1}}}  = \log\frac{|\mathbf{I}_{N_u} + \frac{1}{\sigma_n^2}\mathbf{H}'^H\mathbf{H}'|} {|\mathbf{I}_{|\mathcal{S}_{1}^c|} + \frac{1}{\sigma_n^2} \mathbf{H}_{\mathcal{S}_{1}^c}'^H\mathbf{H}_{\mathcal{S}^c_{1}}'|}, \\
    \qquad\qquad\vdots \\
{R_{{k_{N_u-1}}}} = \log\frac{|\mathbf{I}_{|\mathcal{S}_{N_u-2}^c|} + \frac{1}{\sigma_n^2}\mathbf{H}_{\mathcal{S}^c_{N_u-2}}'^H \mathbf{H}_{\mathcal{S}^c_{N_u-2}}'|}{|{1 + \frac{1}{{\sigma _n^2}}{\mathbf{h}'}_{{k_{N_u}}}^H{{\mathbf{h}'}_{{k_{N_u}}}}}|}, \\
{R_{{k_{{N_u}}}}}  = \log \big( {1 + \frac{1}{{\sigma _n^2}}{\mathbf{h}'}_{{k_{N_u}}}^H{{\mathbf{h}'}_{{k_{N_u}}}}} \big),\\
\end{array} \right.
\end{equation}
\emph{where $(k_1, \cdots,k_{N_u})$ is a permutation of $(1,2,\cdots,N_u)$, $\mathcal{S}_i=\{k_1,\cdots,k_i\}$ for $i=1,\dots,N_u-1$.}
\section{Proof of Corollary 3}\label{APP_Coro1}
For any maximal extreme point expressed in (\ref{dominate}) with order vector [$k_1,\cdots,k_{N_u}]$, we let $\gamma_{k_i}/\gamma_{k_{i-1}}\to \infty$, for any $i\in \mathcal{N}_u/\{1\}$. Therefore, similar to the green curves showed in Fig. \ref{f3} and Fig. \ref{f4}, the user $k_{N_u}$ is recovered after all the variances of other users already being zeros as $\gamma_{k_{N_u}}/\gamma_{k_{i-1}}\to \infty$, for any $i\in \mathcal{N}_u/\{1\}$. Thus, from (\ref{lemma1}), the rate of user $k_{N_u}$ is\vspace{-0.2cm}
\begin{equation}\label{coro1}
R_{k_{N_u}}= \log \big( {1 + \frac{1}{{\sigma _n^2}}{\mathbf{h}'}_{{k_{N_u}}}^H{{\mathbf{h}'}_{{k_{N_u}}}}} \big),
\end{equation}
which is the same as that in (\ref{dominate}).
Similarly, when we recovering user $k_{N_u-1}$, all the users have been recovered except user $k_{N_u}$ and user $k_{N_u}-1$. Hence, based on \emph{Theorem 1}, we have
\begin{equation}\label{coro2}
R_{k_{N_u-1}} \!+\! R_{k_{N_u}} \!\!\!= \! \log|\mathbf{I}_{|\mathcal{S}_{N_u-2}^c|} \!+\! \frac{1}{\sigma_n^2} \mathbf{H}_{\mathcal{S}^c_{N_u\!-\!2}}'^H \mathbf{H}_{\mathcal{S}^c_{N_u\!-\!2}}'|.
\end{equation}
Thus, the rate of user $k_{N_u-1}$ is
\begin{equation}
R_{k_{N_u\!-\!1}}\!\!=\! \log\frac{|\mathbf{I}_{|\mathcal{S}_{N_u-2}^c|} + \frac{1}{\sigma_n^2} \mathbf{H}_{\mathcal{S}^c_{N_u-2}}'^H \mathbf{H}_{\mathcal{S}^c_{N_u-2}}'|}{{1 + \frac{1}{{\sigma _n^2}}{\mathbf{h}'}_{{k_{N_u}}}^H{{\mathbf{h}'}_{{k_{N_u}}}}}},
\end{equation}
which is the same as that in (\ref{dominate}). Continue this process and we can show all the other users' rates are the same as that of in (\ref{dominate}). Therefore, we have \emph{Corollary 3}.

\section*{Acknowledgement}
The authors would like to thank Prof. Li Ping for fruitful discussions.

\end{document}